%% file: sgp.tex
\documentclass[12pt]{article}

\usepackage[top=1in, bottom=1in, left=1in, right=1in]{geometry} 
\usepackage{setspace} 
\usepackage{natbib} 
\usepackage{lineno} 
\usepackage{hyperref} 
\usepackage{graphicx}%
\usepackage{multirow}%
\usepackage{amsmath,amssymb,amsfonts}%
\usepackage{amsthm}%
\usepackage{mathrsfs}%
\usepackage[title]{appendix}%
\usepackage[dvipsnames]{xcolor}%
\usepackage{textcomp}%
\usepackage{manyfoot}%
\usepackage{booktabs}%
\usepackage{algorithm}%
\usepackage{algorithmicx}%
\usepackage{algpseudocode}%
\usepackage{listings}%
\usepackage{natbib}%
\usepackage{tikz} 
\usepackage{tikz-network} 
\usepackage{longtable} 

\newcommand{\bin}{\mathrm{Bin}}
\newcommand{\wei}{\mathrm{Wei}}
\newcommand{\gp}{\mathrm{GP}}

\newcommand{\dgamma}{\mathrm{Ga}}

\newcommand{\igamma}{\mathrm{IGa}}
\newcommand{\eps}{\varepsilon}
\newcommand{\cg}{\mathrm{CoGa}}
\newcommand{\stu}{\mathrm{T}}
\newcommand{\logn}{{\log}N}

\title{A Semi-Parametric Bayesian Spatial Model for Rainfall Events in Geographically Complex Domains}
\author{
	Paolo Onorati\textsuperscript{1} and Antonio Canale\textsuperscript{1} \\
	\small{\textsuperscript{1}Department of Statistical Sciences, University of Padova, Italy}
}
\date{}

\begin{document}

\maketitle

\begin{abstract}
	Environmental phenomena are influenced by complex interactions among various factors. For instance, the amount of rainfall measured at different stations within a given area is shaped by atmospheric conditions, orography, and physics of water processes. Motivated by the need to analyze rainfall across complex spatial locations, we propose a flexible Bayesian semi-parametric model for spatially distributed data. This method effectively accounts for spatial correlation while incorporating dependencies on geographical characteristics in a highly flexible manner. Indeed, using latent Gaussian processes, indexed by spatial coordinates and topographical features, the model integrates spatial dependencies and environmental characteristics within a nonparametric framework. Posterior inference is conducted using an efficient rejection-free Markov Chain Monte Carlo algorithm, which eliminates the need for tuning parameter calibration, ensuring smoother and more reliable estimation. The model's flexibility is evaluated through a series of simulation studies, involving different rainfall and spatial correlation scenarios, to demonstrate its robustness across various conditions. We then apply the model to a large dataset of rainfall events collected from the Italian regions of Veneto and Trentino-Alto Adige, these areas are known for their complex orography and diverse meteorological drivers. By analyzing this data, we generate detailed maps that illustrate the mean and variance of rainfall and rainy days. The method is implemented in a new R package available on GitHub.
\end{abstract}

\begin{center}
\end{center}

\section{Introduction} \label{sec:intro}

Rainfall is a fundamental component of the Earth's hydrological cycle, playing a central role in shaping ecosystems and supporting human activities such as agriculture, water management, and urban planning. Accurate rainfall modeling is crucial for addressing challenges like optimizing water resource allocation, ensuring food security, and  mitigating the impacts of water-related natural disasters, which are often exacerbated in geographically complex regions. Therefore, modeling rainfall is an important topic in environmental sciences and spatial statistics. 

Rain magnitude is typically modelled  under a Weibull model assumption.
The Weibull distribution offers a robust framework for representing rainfall due to both physical motivations and empirical evidence, particularly with regard to tail properties \citep{wilson2005, marra2023}. While the primary focus here is on rainfall, the Weibull distribution has been widely applied to other environmental phenomena such as wind speed \citep{justus1976, justus1978, seguro2000, weisser2003, ramirez2005, gryning2014, harris2014}, temperature \citep{jandhyala1999}, intervals between environmental events \citep{sengupta2015}, and spring discharge \citep{leone2021}. Additionally, its use extends to other fields dealing with a wide range of heavy tailed phenomena, including  biological extinction events or financial market returns \citep{laherrere1998}.

In this paper we are particularly motivated by the analysis of rainfall data in geographically complex regions, where precipitation is shaped by a range of interacting environmental factors. Consistent with this, the literature proposes the use of  nonparametric tools such as Gaussian processes \citep[GP,][]{rasmussen2006}. GPs, widely recognized in geostatistics as \textit{Kriging} \citep{krige1951}, provide a robust and flexible framework for regression.
Their use enables the specification of priors directly in function space under a Bayesian framework \citep{ohagan1978} without the need to resort to any  parametric assumption. This justifies their widely use in many different fields \citep{matheron1973, choi2011, datta2016, schulz2018, levy2018, deringer2021}. In spatial statistics, GPs are particularly well-suited for spatial interpolation tasks, such as producing maps quantifying the distribution of temperature or precipitation \citep{curci2021}.

Many applications of GP, however,  suffer of underestimation of posterior uncertainty. Indeed, a GP is fully defined by a mean function and a covariance function (also named kernel function). While the former typically assumes a linear or constant function whose parameters are estimated in a full Bayesian analysis, for the latter there are available multiple parametric options \citep[Chapter~4]{rasmussen2006} depending from hyper-parameters. However, for computational reasons, these hyper-parameters are often assumed known, or, fixed using ad hoc procedures such as  by maximizing the marginal density of the data. Nevertheless, assuming a known covariance function do not take account the posterior uncertainty regarding the hyper-parameters and thus it may cause a general overestimation of the accuracy \citep{handcock1993, moyeed2002, helbert2009, song2015}. 

A possible way to deal with this phenomenon is the use of the so-called Integrated Nested Laplace Approximation (INLA) methodology that produces an approximation of a fully Bayesian analysis and retains a cheap computational cost \citep{rue2009, martins2013}, as recently done by \citet{fioravanti2023} for daily minimum and maximum temperature.

The primary aim of this paper is hence twofold: first, to develop a general and flexible fully Bayesian semi-parametric model that leverages latent GPs to analyze spatially distributed Weibull environmental variables in complex geographical areas; and second, to introduce an accessible and efficient computational approach for posterior inference based on the elliptical slice sampling of  \citet{murray2010}.

The rest of this paper is organized as follows. In Section \ref{sec:model} a Binomial-Weibull  model with latent GP for the parameters and their prior distributions are introduced. Section \ref{sec:computation} describes an efficient computational strategy for Markov Chair Montecarlo (MCMC) posterior inference. In Section \ref{sec:simulation}, the performance of the semi-parametric model are assessed through a simulation study. In Section \ref{sec:real}, the proposed approach is used to model  a rich dataset of rainfall events in North-East of Italy, a region characterized by climatic heterogeneity and complex orography. Finally, in Section \ref{sec:concl}, we present some conclusions.


\section{Model and Prior Distributions} \label{sec:model}

Let $S = \{ S_1 \times S_2 \dots \times S_p \}$ be a $p$-dimensional domain that combines spatial coordinates and topographical features. For $j = 1, 2, \dots T$, assume to model the number of wet days in the generic year $j$ and $s \in S$, as a binomial random variable, i. e. 
\begin{equation}
	N_{j} \vert \pi_j(s) \sim \bin \Big(n = 365, p(s) = (1 + e^{-\pi_j(s)})^{-1} \Big),
 \label{eq:binomial}
\end{equation}
where the probability of rain in a given day $p(s)$, is a generic function of $s$ through $\pi_j(s)$. Conditionally on the total number of wet days $N_j(s)$, assume that rainfall follows a two-parameters Weibull distribution. Notably the choice of Weibull distribution is a standard to describe the distribution of precipitation events \citep{wilson2005}. Specifically, said, $W_{i,j}(s)$ is the magnitude for the $i$-th event at spatial point $s$ in the $j$-th year, we let for $i = 1, 2, \dots, N_j$
\begin{equation}
	W_{i,j}(s) \vert \big( \gamma_{j}(s), \delta_{j}(s) \big) \sim  \wei \Big( \exp \big( \gamma_{j}(s) \big), \exp \big( \delta_{j}(s) \big) \Big),
 \label{eq:weibul}
\end{equation} 
where, consistently with \eqref{eq:binomial}, both parameters are function of $s \in S$.

We model the parameters of the binomial and Weibull distributions in a nonparametric way using noised independent GPs; that is,
\begin{equation*}
	\pi_j(s) = \mu_\pi(s) + \eps_{\pi, j}(s), \quad 
	\gamma_j(s) = \mu_\gamma(s) + \eps_{\gamma, j}(s), \quad
	\delta_j(s) = \mu_\delta(s) + \eps_{\delta, j}(s),
\end{equation*} 
with
\begin{equation}
    \label{eq:GPs}
    \mu_\pi(s) \sim \gp \big( \psi_\pi, k_\pi(s, s^\prime) \big) \, , \,
	\mu_\gamma(s) \sim \gp \big( \psi_\gamma, k_\gamma(s, s^\prime) \big) \, , \,
	\mu_\delta(s) \sim \gp \big( \psi_\delta, k_\delta(s, s^\prime) \big),
\end{equation}
and independent idyosincratic errors
\begin{equation*}
    \eps_{j, \pi}(s) \overset{i.i.d.}{\sim} N(0, \tau^2_\pi) \, , \;
    \eps_{j, \gamma}(s) \overset{i.i.d.}{\sim} N(0, \tau^2_\gamma) \, , \;
    \eps_{j, \delta}(s) \overset{i.i.d.}{\sim} N(0, \tau^2_\delta) \, .
\end{equation*}
For all processes, we use a covariance function based on \textit{Mat\'ern kernels} with shape parameter $\nu = 3/2$, as commonly done in spatial statistics applications with latent GPs \citep{rue2009}. However, here we use a product of kernels, say 
\begin{equation}
	\label{eq:kernel}
	k(s, s') = \sigma^2 \exp \left( -\sqrt{3} \sum_{h=1}^{p} \frac{\vert s_h - s'_h \vert}{\lambda_h} \right) \prod_{h=1}^{p} \left( 1 + \sqrt{3} \: \frac{\vert s_h - s'_h \vert}{\lambda_h} \right).
\end{equation}
From \eqref{eq:kernel} and \eqref{eq:GPs}, it is apparent that the total number of hyper-parameters is $3 (p + 3)$, that is, $(\psi_\pi, \tau^2_\pi, \sigma^2_\pi, \lambda_{\pi, 1}, \lambda_{\pi, 2}, \dots, \lambda_{\pi, p})$, $(\psi_\gamma, \tau^2_\gamma, \sigma^2_\gamma, \lambda_{\gamma, 1}, \lambda_{\gamma, 2}, \dots, \lambda_{\gamma, p})$, and $(\psi_\delta, \tau^2_\delta, \sigma^2_\delta, \lambda_{\delta, 1}, \lambda_{\delta, 2}, \dots, \lambda_{\delta, p})$.

\begin{figure}
	\centering
	\includegraphics[scale=0.75]{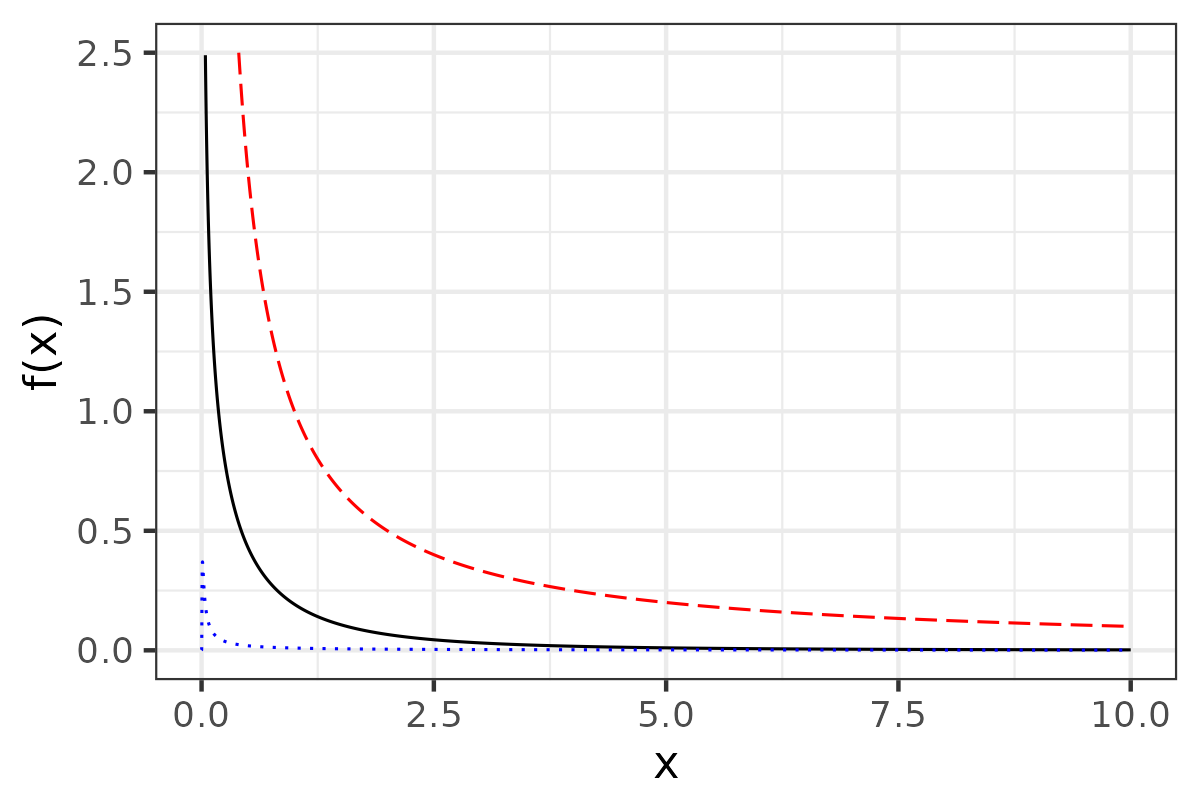}
	\caption{Densities of $\cg(0.5, 2, 2)$ (continuous) vs $\igamma(0.01, 0.01)$ (dotted) vs scale-invariant (long dashed).}
	\label{fig:icg_ig_logunif_densities}
\end{figure}

The following prior distributions are assumed
\begin{align}
    \psi_\pi \sim \stu_2 \, &, \, \tau^2_\pi \sim \cg \left( \frac{1}{2}, 2, 2 \right) \, , \notag \\
    \sigma^2_\pi \sim \cg \left( \frac{1}{2}, 2, 2 \right) \, &, \, \lambda_{\pi, h} \sim \logn(0, 2).
    \label{eq:hyperprior}
\end{align}
Where, $\stu_\nu$ denotes a Student-$t$ distribution with $\nu$ degrees of freedom, $\logn (\mu, \sigma^2)$ denotes a log-normal distribution with log-mean $\mu$ and log-variance $\sigma^2$, and $\cg(k, v, r)$ denotes a compounded gamma distribution, that is, if $X \vert Z \sim \igamma(k, Z)$ and $Z \sim \dgamma(v, r)$, then $X \sim \cg(v, k, 1/r)$.  In the next section, we will exploit the following  stochastic representations
\begin{align*}
	\psi_\pi \vert \zeta_{\psi, \pi} \sim N(0, \zeta_{\psi, \pi}) \, &, \: \zeta_{\psi, \pi} \sim \igamma \left( \frac{1}{2}, \frac{1}{2} \right), \\
	\tau^2_\pi \vert \zeta_{\tau^2, \pi} \sim \igamma \left( 2, \zeta_{\tau^2, \pi} \right) \, &, \: \zeta_{\tau^2, \pi} \sim \dgamma \left( \frac{1}{2}, \frac{1}{2} \right), \\
	\sigma^2_\pi \vert \zeta_{\sigma^2, \pi} \sim \igamma \left( 2, \zeta_{\sigma^2, \pi} \right) \, &, \: \zeta_{\sigma^2, \pi} \sim \dgamma \left( \frac{1}{2}, \frac{1}{2} \right),
\end{align*} 
and similarly for $\psi_\gamma, \psi_\delta, \tau^2_\gamma, \tau^2_\delta, \sigma^2_\gamma, \sigma^2_\delta$. 
 
Notably, the prior distributions for the location and scale parameters promotes sparsity. Indeed, a Student-$t$ distribution with $2$ degrees of freedom has an infinite variance. The density of the prior for the scale parameters resembles the scale-invariant improper prior proportional to the reciprocal of the argument. In fact, our prior choice for scale parameters is unbounded close to $0$ and decreases monotonically. Furthermore, the tail is heavy because, for an $\cg(0.5, 2, 2)$ distribution, the variance is infinite. Figure \ref{fig:icg_ig_logunif_densities} compares the density function of an inverse compounded gamma, an inverse gamma, and the scale invariant improper prior. The priors for the hyper-parameters of $\gamma$ and $\delta$ are analogous.

\begin{figure}[ht]
	\centering
	\input{./figures/dag.tex}
	\caption{Graphical representation of the Bayesian semiparametric model.}
	\label{fig:dag_model}
\end{figure}
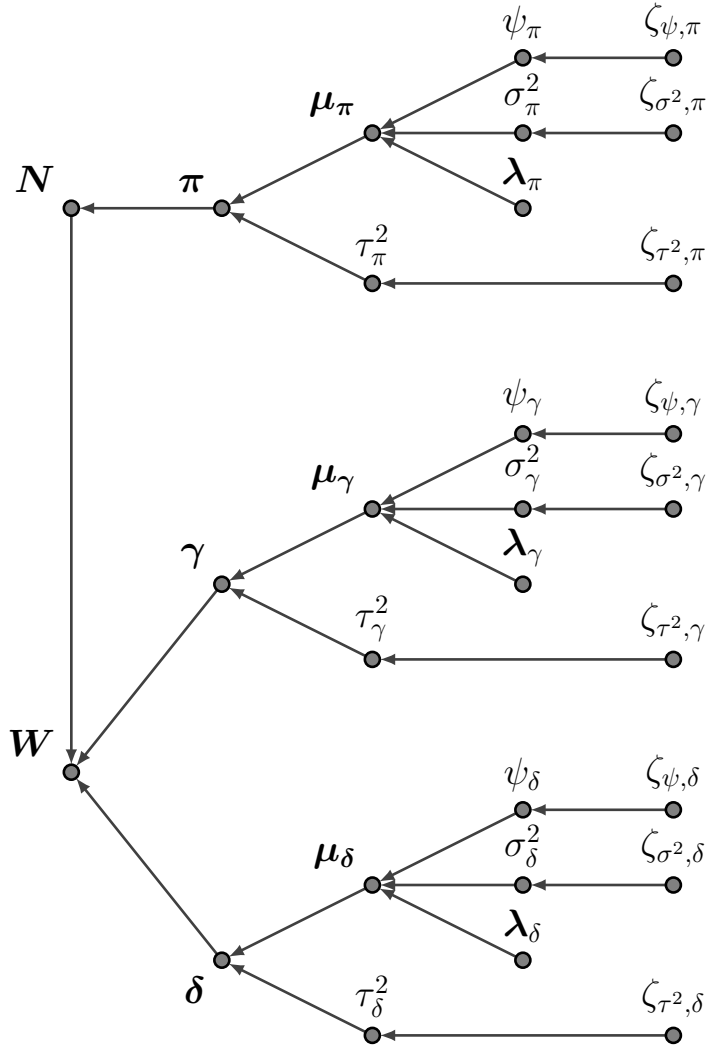

We called the model described in \eqref{eq:binomial}-\eqref{eq:weibul},  equipped with the hierarchical prior distribution discussed in \eqref{eq:GPs}-\eqref{eq:hyperprior}, \textit{semi-parametric model} because the number and magnitude of events is modeled in a parametric way using the binomial and Weibull distributions, while the parameters of these random variables are connected to covariates in a nonparametric way using latent GPs. A pictorial representation of the hierarchical model is represented in Figure \ref{fig:dag_model}.

\section{Posterior Computation} \label{sec:computation}

The full posterior distribution of all the unknowns does not lend itself to a straightforward closed-form expression. Consequently, we employ a MCMC approach to generate samples from the posterior distribution of all parameters. Specifically, we propose to use the elliptical slice sampling (ESS) of \citet{murray2010}. The ESS method shares many similarities with the Metropolis-Hasting algorithm. However, at every iteration, the ESS provides an automatic way to tune the proposal in order to avoid a rejection. Therefore, this method is a rejection-free algorithm, that is, the probability that $2$ consecutive draws are equal is $0$. For using this approach, the target distribution to be sampled must be expressed as a product between a centered Gaussian density and a positive function, see \citet{murray2010} for  details.

Conditionally on the observation of data for $M$ distinct spatial points, with  $\boldsymbol{s} = s_1, s_2, \dots, s_M$, let $\boldsymbol{N} = \{ N_j(s_m): j = 1, 2, \dots, T \; , \; m = 1, 2, \dots, M \}$ be the column-vector of all observed binomial draws in the $M$ spatial points at the times $t = 1, 2, \dots, T$. Similarly, we define the following column-vectors, 
\begin{align*}
	\boldsymbol{W} &= \{ W_{i,j}(s_m): i = 1, 2, \dots, N_j(s_m) \; , \; j = 1, 2, \dots, T \; , \; m = 1, 2, \dots, M \} \, , \\
	\boldsymbol{\pi} &= \{ \pi_j(s_m): j = 1, 2, \dots, T \; , \; m = 1, 2, \dots, M \} \, , \\
	\boldsymbol{\gamma} &= \{ \gamma_{i,j}(s_m): i = 1, 2, \dots N_j(s_m) \; , \; j = 1, 2, \dots, T \; , \; m = 1, 2, \dots, M \} \, , \\
	\boldsymbol{\delta} &= \{ \delta_{i,j}(s_m): i = 1, 2, \dots N_j(s_m) \; , \; j = 1, 2, \dots, T \; , \; m = 1, 2, \dots, M \}, \\
	\boldsymbol{\mu_\pi} &= \{ \mu_\pi(s_m): m = 1, 2, \dots, M \}, \\
	\boldsymbol{\mu_\gamma} &= \{ \mu_\gamma(s_m): m = 1, 2, \dots, M \}, \\
	\boldsymbol{\mu_\delta} &= \{ \mu_\delta(s_m): m = 1, 2, \dots, M \}, \\
	\boldsymbol{\lambda_\pi} &= \{ \lambda_{\pi, h}: h = 1, 2, \dots, p \}, \\
	\boldsymbol{\lambda_\gamma} &= \{ \lambda_{\gamma, h}: h = 1, 2, \dots, p \}, \\
	\boldsymbol{\lambda_\delta} &= \{ \lambda_{\delta, h}: h = 1, 2, \dots, p \}.
\end{align*}
Furthermore, let $K_\pi(\boldsymbol{s}, \boldsymbol{s}^\prime \vert \sigma^2_\pi, \boldsymbol{\lambda_\pi})$ be the variance-covariance matrix of $\boldsymbol{\mu_\pi}$. Here, the notation $K_\pi(\cdot, \cdot \vert \sigma_\pi^{2}, \boldsymbol{\lambda_\pi})$ means that we are using the Mat\'ern kernel \eqref{eq:kernel} with hyper-parameter values $\sigma^2_\pi$ and $\boldsymbol{\lambda_\pi}$. We use the same notation, say $K_\gamma(\boldsymbol{s}, \boldsymbol{s}^\prime \vert \sigma_\gamma^{2, (k+1)}, \boldsymbol{\lambda_\gamma}^{(k+1)})$ and $K_\delta(\boldsymbol{s}, \boldsymbol{s}^\prime \vert \sigma_\delta^{2, (k+1)}, \boldsymbol{\lambda_\delta}^{(k+1)})$, for the GPs modeling the parameters of Weibull distributions. Also, we denote $1_M$ the column-vector of $M$ ones.

Notice that, conditioned on $\boldsymbol{N}$, inference on the parameters of the binomial distributions is independent from that on the parameters of the Weibull distributions. This implies that, if we consider $\boldsymbol{N}$ constant, we can completely remove the latent variables related to binomial parameters and skip their updates in the sampling algorithm; this does not affect the inference for the Weibull distributions. This simplified approach is used for the simulation study presented in Section \ref{sec:simulation}. However, in the application discussed in Section \ref{sec:real} we make inference for all the parameters of the model.

Denoting the values of the parameters at $k$-th iteration with upper round parenthesis, the update of the parameters iterates the following steps:

\begin{enumerate}
	\item Update scaling variable of stochastic representations of Student-$t$ and inverse compounded gamma distributions:
				\vspace{2.5mm}
				\begin{enumerate}
					\item Sample $\zeta^{(k+1)}_{\psi, \pi} \vert \psi^{(k)}_\pi$ from
								$
									\igamma \left( \frac{3}{2}, 1 + \frac{\psi^{2,(k)}_\pi}{2} \right) \, ,
								$
					
					\item Sample $\zeta^{(k+1)}_{\tau^2, \pi} \vert \tau^{2, (k)}_\pi$ from
								$
									\dgamma \left( \frac{5}{2}, \frac{1}{2} + \frac{1}{\tau^{2, (k)}_\pi} \right);$
					
					\item Sample $\zeta^{(k+1)}_{\sigma^2, \pi} \vert \sigma^{2, (k)}_\pi$ from 
								$
									\dgamma \left( \frac{5}{2}, \frac{1}{2} + \frac{1}{\sigma^{2, (k)}_\pi} \right);$
					\item Proceed similarly for $\zeta^{(k+1)}_{\psi, \gamma}, \zeta^{(k+1)}_{\tau^2, \gamma}, \zeta^{(k+1)}_{\sigma^2, \gamma}$ and $\zeta^{(k+1)}_{\psi, \delta}, \zeta^{(k+1)}_{\tau^2, \delta}, \zeta^{(k+1)}_{\sigma^2, \delta}$.
				\end{enumerate}
	
	\vspace{2.5mm}
	\item Update variances of white noises and variances of GPs:
				\vspace{2.5mm}
				\begin{enumerate}
					\item Sample $\tau^{2, (k+1)}_\pi \vert \zeta^{(k+1)}_{\tau^2, \pi}, \psi^{(k)}_{\pi}, \boldsymbol{\mu_\pi}^{(k)}$ from
								\begin{equation*}
									\igamma \left( 2 + \frac{M}{2}, \zeta^{(k+1)}_{\tau^2, \pi} + \frac{\sum_{m = 1}^{M} \left( \psi^{(k)}_\pi - \mu^{(k)}_\pi(s_m) \right)^2}{2} \right) \, , \\
								\end{equation*}

					\item Sample $\sigma^{2, (k+1)}_\pi \vert \zeta^{(k+1)}_{\sigma^2, \pi}, \boldsymbol{\mu_\pi}^{(k)}, \boldsymbol{\pi}^{(k)}$ from
								\begin{equation*}
									\igamma \left( 2 + \frac{M \, T}{2}, \zeta^{(k+1)}_{\sigma^2, \pi} + \frac{\sum_{m = 1}^{M} \sum_{j = 1}^{T} \left( \mu^{(k)}_\pi(s_m) - \pi^{(k)}_j(s_m) \right)^2}{2} \right) \, , \\
								\end{equation*}
					\item Proceed similarly for  $\tau^{2, (k+1)}_\gamma, \sigma^{2, (k+1)}_\gamma$ and $\tau^{2, (k+1)}_\delta, \sigma^{2, (k+1)}_\delta$.
				\end{enumerate}
	
	\vspace{2.5mm}
	\item Update length scales of GPs:
				\vspace{2.5mm}
				\begin{enumerate}
					\item The density of $\boldsymbol{\lambda_\pi}^{(k+1)} \vert \sigma^{2, (k+1)}_\pi, \boldsymbol{\mu_\pi}^{(k)}, \psi^{(k)}_\pi$ is propotional to
								\begin{align*}
									 f & \left( \boldsymbol{\mu_\pi}^{(k)} \vert \sigma^{2, (k+1)}_\pi, \boldsymbol{\lambda_\pi}^{(k+1)}, \psi^{(k)}_\pi \right) f \left( \boldsymbol{\lambda_\pi}^{(k+1)} \right) \, ,
								\end{align*}
					and the second factor is a product of log-normal densities with logarithmic mean $0$. Thus, it is possible to use an ESS step on the log-scale for updating $\boldsymbol{\lambda_\pi}^{(k+1)}$ as reported in the Supplementary Materials. 
				  
					\vspace{2.5mm}
					\item Proceed similarly for  $\boldsymbol{\lambda_\gamma}^{(k+1)}$ and $\boldsymbol{\lambda_\delta}^{(k+1)}$.
				\end{enumerate}

	\vspace{2.5mm}
	\item Update mean functions, GPs and parameters:
	\vspace{2.5mm}
	\begin{enumerate}
		\item The density of $\psi^{(k+1)}_\pi, \boldsymbol{\mu_\pi}^{(k+1)}, \boldsymbol{\pi}^{(k+1)} \vert \boldsymbol{N}, \tau^{2, (k+1)}_\pi, \sigma^{2, (k+1)}_\pi, \boldsymbol{\lambda_\pi}^{(k+1)}, \zeta^{k+1}_{\psi, \pi}$ is propotional to
					\begin{equation*}
						 f \left( \boldsymbol{N} \vert \boldsymbol{\pi}^{(k+1)} \right) f \left( \boldsymbol{\pi}^{(k+1)}, \boldsymbol{\mu_\pi}^{(k+1)}, \psi^{(k+1)}_\pi \vert \tau^{2, (k+1)}_\pi, \sigma^{2, (k+1)}_\pi, \boldsymbol{\lambda_\pi}^{(k+1)}, \zeta^{k+1}_{\psi, \pi} \right) \, ,
					\end{equation*}
					and the second factor is a joint Gaussian density with mean equal to the null vector, thus it is possible to use an ESS step for updating $\psi^{(k+1)}_\pi, \boldsymbol{\mu_\pi}^{(k+1)}, \boldsymbol{\pi}^{(k+1)}$ as reported in the Supplementary Materials.
		
		\vspace{2.5mm}
		\item The density of $\psi^{(k+1)}_\gamma, \boldsymbol{\mu_\gamma}^{(k+1)}, \boldsymbol{\gamma}^{(k+1)}, \psi^{(k+1)}_\delta, \boldsymbol{\mu_\delta}^{(k+1)}, \boldsymbol{\delta}^{(k+1)} \vert \boldsymbol{W}, \boldsymbol{N}, $ $ \tau^{2, (k+1)}_\gamma, \sigma^{2, (k+1)}_\gamma, \boldsymbol{\lambda_\gamma}^{(k+1)}, \zeta^{k+1}_{\psi, \gamma}, \tau^{2, (k+1)}_\delta, \sigma^{2, (k+1)}_\delta, \boldsymbol{\lambda_\delta}^{(k+1)}, \zeta^{k+1}_{\psi, \delta}$ is proportional to
					\begin{align*}
						&f \left( \boldsymbol{W} \vert \boldsymbol{N}, \boldsymbol{\gamma}^{(k+1)}, \boldsymbol{\delta}^{(k+1)} \right) \times \\
						&\times f \left( \boldsymbol{\gamma}^{(k+1)}, \boldsymbol{\mu_\gamma}^{(k+1)}, \psi^{(k+1)}_\gamma \vert \tau^{2, (k+1)}_\gamma, \sigma^{2, (k+1)}_\gamma, \boldsymbol{\lambda_\gamma}^{(k+1)}, \zeta^{k+1}_{\psi, \gamma} \right) \times \\
						&\times f \left( \boldsymbol{\delta}^{(k+1)}, \boldsymbol{\mu_\delta}^{(k+1)}, \psi^{(k+1)}_\delta \vert \tau^{2, (k+1)}_\delta, \sigma^{2, (k+1)}_\delta, \boldsymbol{\lambda_\delta}^{(k+1)}, \zeta^{k+1}_{\psi, \delta} \right) \, ,
					\end{align*}
		and the latter $2$ factors are a product between $2$ joint Gaussian densities with both means equal to the null vector, thus it is possible to use an ESS step for updating $\psi^{(k+1)}_\gamma, \boldsymbol{\mu_\gamma}^{(k+1)}, \boldsymbol{\gamma}^{(k+1)}$ and $\psi^{(k+1)}_\delta, \boldsymbol{\mu_\delta}^{(k+1)}, \boldsymbol{\delta}^{(k+1)}$ as reported in the Supplementary Materials
	\end{enumerate}
	 
\end{enumerate}
In steps $1$ and $2$, we are able to sample directly from the full conditionals. In steps $3$ and $4$, we do not have a simple distribution but we can use an ESS step. This implies that, at every update, we obtain a value that is different from the previous one and then the entire algorithm is rejection-free. In addition, using elliptical slice sampling is easier than using a Metropolis-Hasting approach because it does not require to the user to select any tuning parameters.

The algorithms described in this section, and in the related subsections, are implemented in the \texttt{spg} R package available on \texttt{BLINDED FOR REVIEW}. 

\subsection{Forecasting} \label{subsec3.1}

The model can be used to forecast the number of events and their magnitude at unobserved points in order to create maps associated to different aspects of the rainfall distribution. In our motivating applications, these points represent pixels of a map.

Let $\boldsymbol{s^\ast} = \{ s^\ast_1, s^\ast_2, \dots, s^\ast_{M^\ast} \}$ be a set of $M^\ast$ points for which we do not observe the number of events and their magnitude. Let $K_\pi(s^\ast_{m^\ast}, \boldsymbol{s}^\prime \vert \sigma^2_\pi, \boldsymbol{\lambda_\pi})$ be the row-vector of cross-covariance of $(\mu_\pi(s^\ast_{m^\ast}), \boldsymbol{\mu_\pi})$. We use the same notation, say $K_\gamma(s^\ast_{m^\ast}, \boldsymbol{s}^\prime \vert \sigma^2_\gamma, \boldsymbol{\lambda_\gamma})$ and $K_\delta(s^\ast_{m^\ast}, \boldsymbol{s}^\prime \vert \sigma^2_\delta, \boldsymbol{\lambda_\delta})$, for the other GPs modeling the parameters of the Weibull distributions.

Hence, at the end of $k$-th iteration of the MCMC algorithm, we can forecast the number of events and their magnitude at $\boldsymbol{s^\ast}$ using the following steps:
\begin{enumerate}
	\item[] For $m^\ast = 1, 2, \dots, M^\ast$:
	 \begin{enumerate}
		\vspace{2.5mm}
		\item[1.] Sample $\mu_\pi^{(k)}(s^\ast_{m^\ast}) \vert \boldsymbol{\mu_\pi}^{(k)}, \psi^{(k)}_\pi, \sigma^{2, (k)}_\pi, \boldsymbol{\lambda_\pi}^{(k)}$ from an univariate Gaussian distribution with mean 
					\begin{equation*}
						\psi^{(k)}_\pi + %
						K_\pi(s^\ast_{m^\ast}, \boldsymbol{s}^\prime \vert \sigma_\pi^{2, (k)}, \boldsymbol{\lambda_\pi}^{(k)}) %
						K^{-1}_\pi(\boldsymbol{s}, \boldsymbol{s}^\prime \vert \sigma_\pi^{2, (k)}, \boldsymbol{\lambda_\pi}^{(k)}) %
						( \boldsymbol{\mu_\pi}^{(k)} - \psi^{(k)}_\pi 1_M ),
      \end{equation*}
      and variance
      \begin{equation*}
						\sigma^{2,(k)}_\pi - %
						K_\pi(s^\ast_{m^\ast}, \boldsymbol{s}^\prime \vert \sigma_\pi^{2, (k)}, \boldsymbol{\lambda_\pi}^{(k)}) %
						K^{-1}_\pi(\boldsymbol{s}, \boldsymbol{s}^\prime \vert \sigma_\pi^{2, (k)}, \boldsymbol{\lambda_\pi}^{(k)}) %
						K^\prime_\pi(s^\ast_{m^\ast}, \boldsymbol{s}^\prime \vert \sigma_\pi^{2, (k)}, \boldsymbol{\lambda_\pi}^{(k)}) \, .
					\end{equation*}
					Do the same for $\mu_\gamma^{(k)}(s^\ast_{m^\ast})$ and $\mu_\delta^{(k)}(s^\ast_{m^\ast})$.

		\vspace{2.5mm}
		\item[2.] Sample $\pi^{(k)}(s^\ast_{m^\ast}) \vert \mu_\pi^{(k)}(s^\ast_{m^\ast}), \tau^{2, (k)}_\pi$ from a
					\begin{align*}
						N \left( \mu_\pi^{(k)}(s^\ast_{m^\ast}), \tau^{2, (k)}_\pi \right) \, .
					\end{align*}
					Do the same for $\gamma^{(k)}(s^\ast_{m^\ast})$ and $\delta^{(k)}(s^\ast_{m^\ast})$.
	 \end{enumerate}
\end{enumerate}
At the end of the above scheme, we can either (i) sample $N^{(k)}(s^\ast_{m^\ast})$ and $W_j^{(k)}(s^\ast_{m^\ast}), \, j = 1, 2, \dots, N^{(k)}(s^\ast_{m^\ast})$, to obtain a posterior sample of any functional of interest, or (ii) compute relevant quantities for which closed-formula expressions in terms of the parameters exist, such as mean, variance, and Kullback-Leibler divergence. In the latter case, we still obtain a posterior sample of the quantities of interest. Whenever we need a Bayesian point estimator, we use the posterior median.

Notably, this approach samples from the posterior marginal distribution of the unobserved points. Thus, independence between them is assumed for the sake of computational cost, indeed, if one samples from the joint distribution, in step $1$ the variance is replaced with an $M^\ast \times M^\ast$ matrix whose Cholesky decomposition must be computed. In the case of drawing a map, this is very useful because the unobserved points are pixels and their number is generally large, and the correctness of the map is preserved if the map represents some quantities from the marginal distribution of the pixels, as usual.

\section{Simulation Study} \label{sec:simulation}

In this section we perform a simulation study to assess the performance of the proposed semiparametric model. As competitor, we consider a model where the parameters of Equations \eqref{eq:binomial} and \eqref{eq:weibul} are simple linear functions of $s \in S$. In the context of spatial extreme rainfall analysis, a similar approach have been used by \citet{stolf2023}. All the technical details, including parameter specification, prior elicitation, and posterior computation of the competing parametric model are reported in the Supplementary Materials. 

For simplicity, we focus exclusively on inference for Weibull distributions, as the estimates of the binomial parameters are independent of those for the Weibull parameters. Furthermore, inference on the model component related solely to binomial variables is a well-known nonparametric classification model, for which extensive literature is available \citep[Chapter~3]{rasmussen2006}.

We consider the case of $p = 2$ covariates with range $(-1, 1)$ mimicking latitude and longitude. We analyze two different scenarios. In the first one the true functions generating $\mu_\gamma(s)$ and $\mu_\delta(s)$ are far from linearity and we denote this as non-linear scenario. In the second case, the true functions are exactly linear. We combine these $2$ scenarios with $2$ different sample sizes, namely $M = 31$ and $M = 64$. The other quantities are kept fixed for all scenarios, in particular, we set $T = 4$, $N_j(s_m) = 134 \, , \, j = 1, 2, \dots T \, , \, m = 1, 2, \dots, M$. For each combination of scenario and sample size, we simulated $R=128$ independent replicates. In the Supplementary Materials, we provide the point locations for both settings. Let $\gamma^\dag(s_m), \delta^\dag(s_m)$ be the true parameters of the Weibull distributions at $s_m$. The exact  expressions generating $\gamma^\dag(s_m)$ and $\delta^\dag(s_m)$ both for the non-linear and linear scenarios (see Figure \ref{fig:true_funs} for their plots), are reported in the Supplementary Materials. 
\begin{figure}
	\centering
	\includegraphics{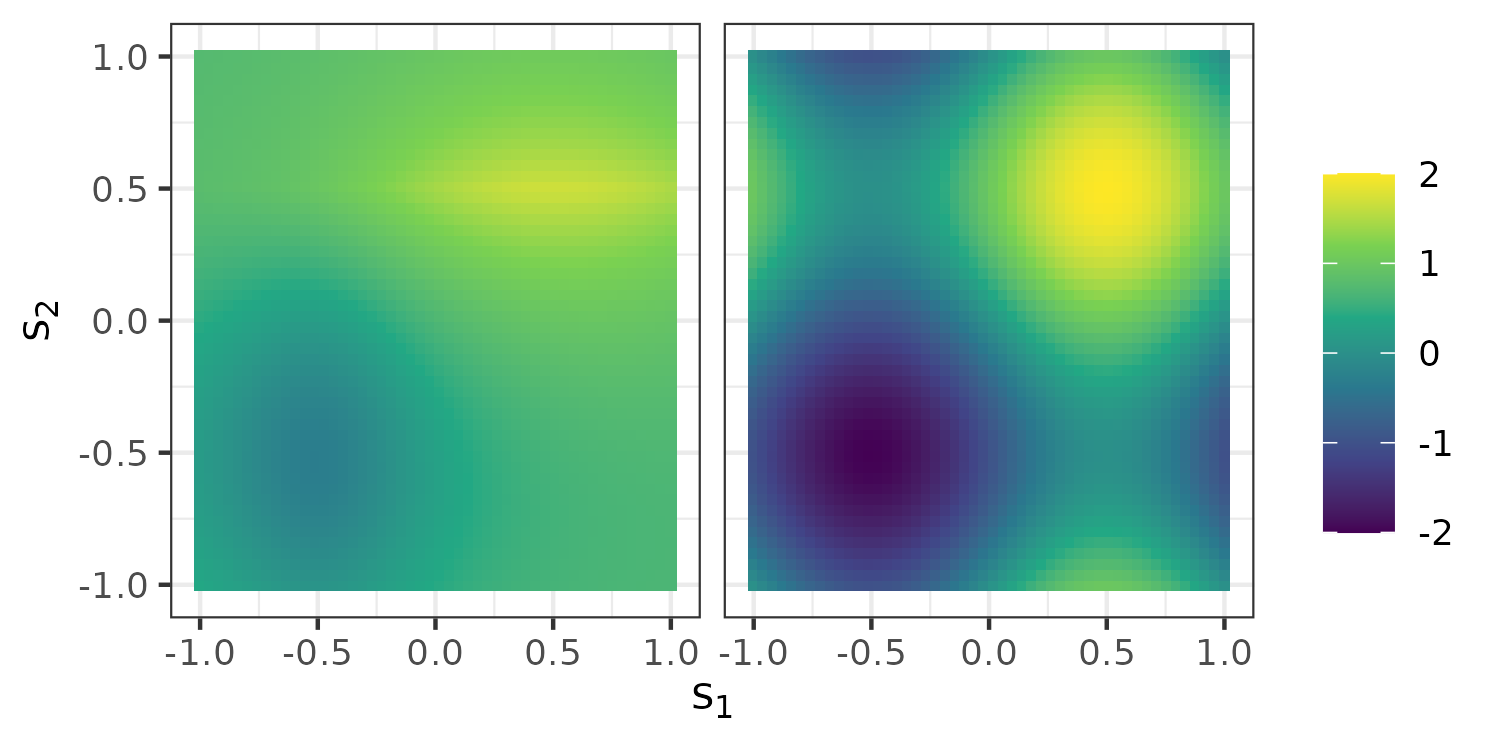}
	\caption{Simulation study: true functions in the non-linear scenario for Weibull parameters: $\gamma^\dag(s)$ (left) and $\delta^\dag(s)$ (right).}
	\label{fig:true_funs}
\end{figure}

We run the proposed MCMC algorithms for semi-parametric and parametric models for $10^4$ iterations. For each iteration, we consider a grid of points in the square defined by the corners $(1, 1$), $(-1, 1)$, $(1, -1)$, and $(-1, -1)$, we forecast the parameters of the Weibull random variables and we compute the Kullback-Leibler divergence from the true distributions. Thus, we end-up with a posterior sample of divergences for each point belonging to the grid for each independent sample.

\begin{figure}
	\centering
	\includegraphics{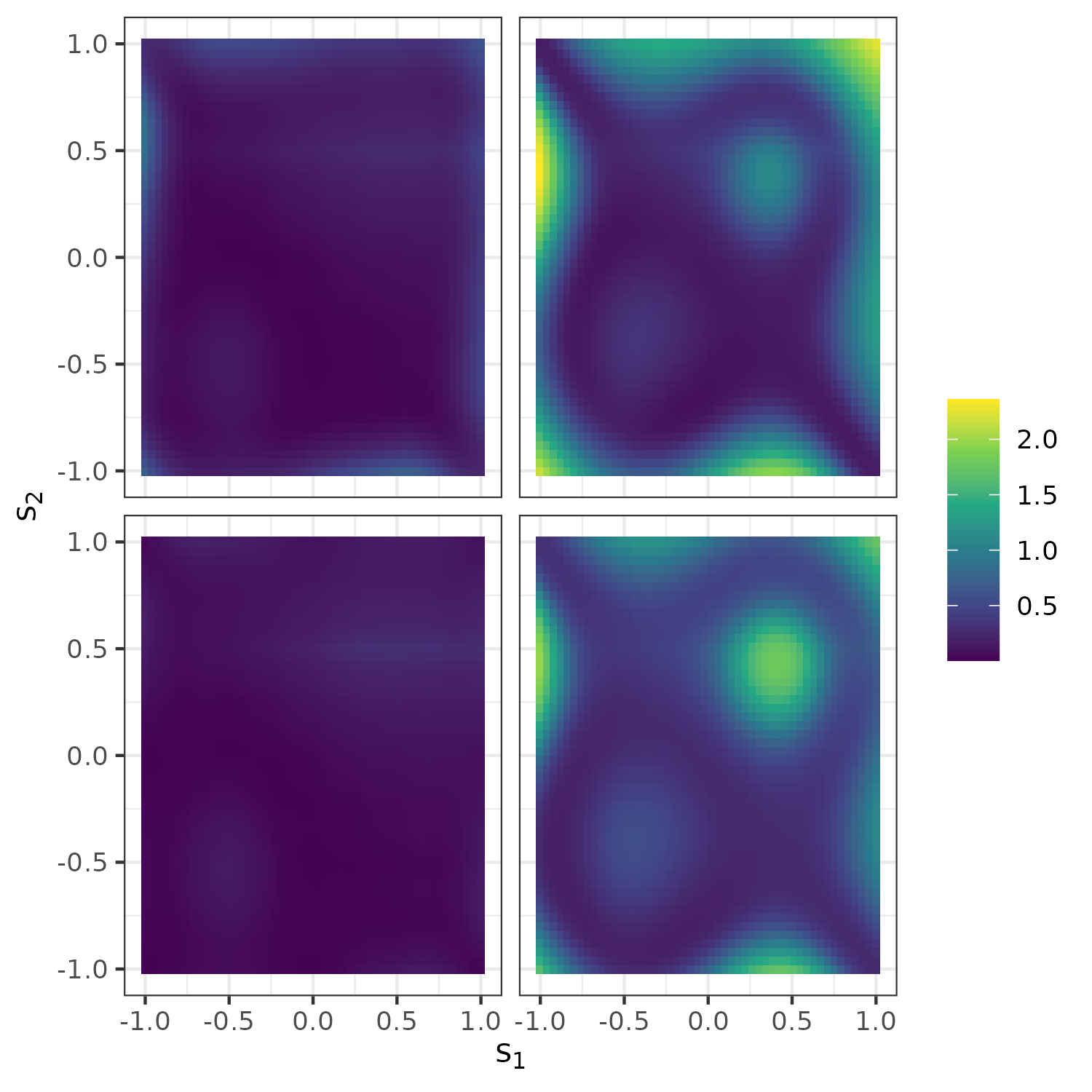}
	\caption{Simulation study: logarithm of sample mean of Kullback-Leibler divergence estimates for the non-linear scenario: semi-parametric (left) vs parametric (right), small sample size (top) and large sample size (bottom).}
	\label{fig:klmaps_nolm}
\end{figure}

\begin{figure}
	\centering
	\includegraphics{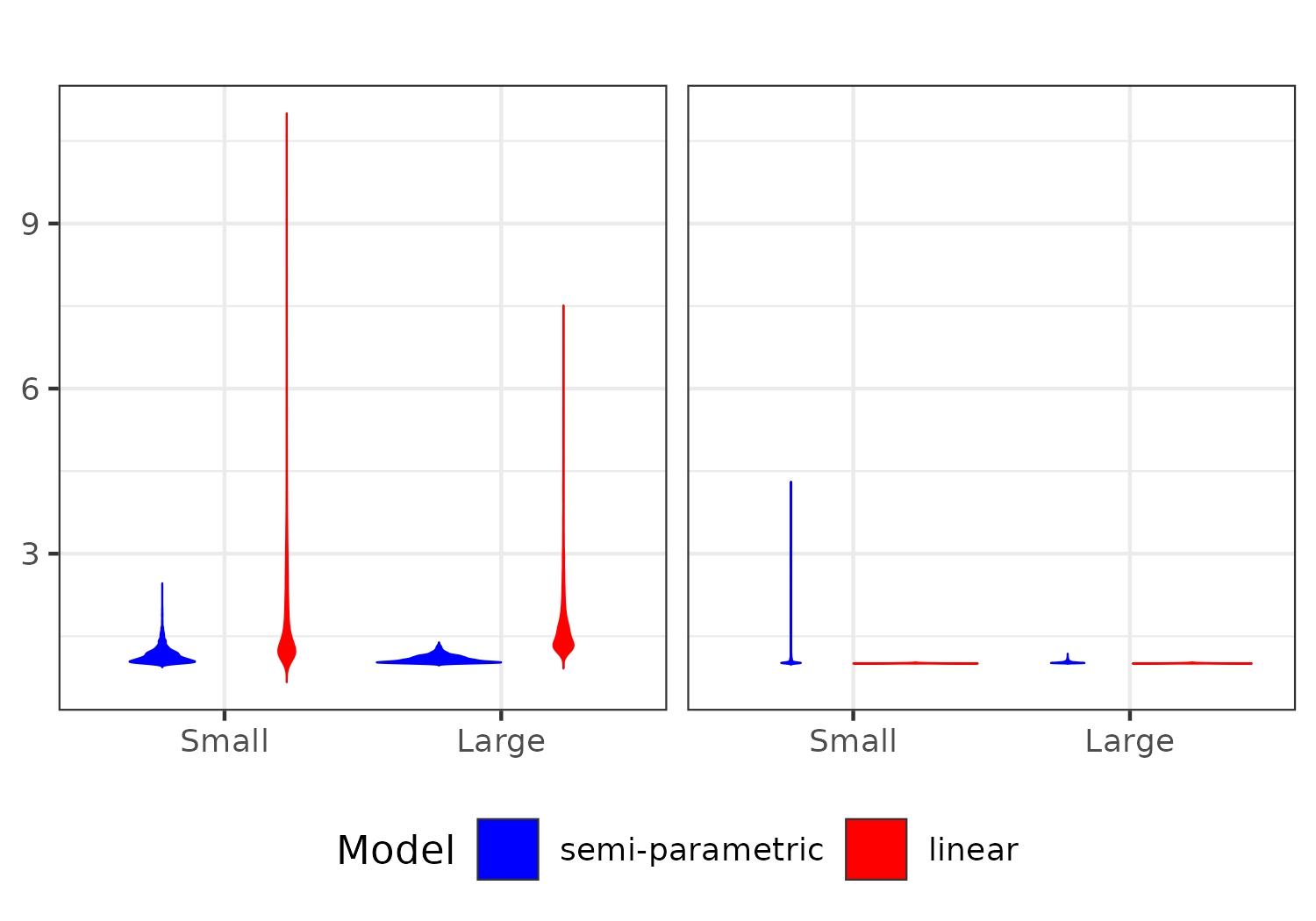}
	\caption{Simulation study: violin plots for sample mean of Kullback-Leibler divergence estimates: non-linear (left) vs linear (right) scenario.}
	\label{fig:klviolin}
\end{figure}

Figure \ref{fig:klmaps_nolm} reports the logarithm of the average posterior medians of the Kullback-Leibler divergences, obtained averaging over the $R$ replicates, for each pixel of the map in the non-linear scenario. Similarly, the first panel of Figure \ref{fig:klviolin} displays the distributions of the same quantities on the true scale using violins. As expected, the results show that the semi-parametric model outperforms the parametric one in the non-linear scenario. Specifically, the semi-parametric model is able to capture the non-linearity in the data and its goodness-of-fit increases as the sample size gets large. On the other hand, the parametric model exhibits a poor performances in the regions where a peak or a valley is present in the true functions. There is not even a clear improvement of the goodness-of-fit with an increase in sample size, confirming the observed low accuracy as a consequence  of the  model's poor flexibility. Figure \ref{fig:klmaps_nolm}, in the Supplementary Materials and the second panel of  Figure \ref{fig:klviolin}, shows the same quantities for the linear scenarios. In this case,  the parametric model offers advantages in both the rate of convergence to the true model as sample size increases and in computational efficiency. Nonetheless, the semi-parametric model still captures the linearity in the data, albeit with reduced efficiency. 



\section{Analysis of the Rainfall in the North-East of Italy} \label{sec:real}

We analyze a dataset of rainfall in an Alpine region spanning approximately 32,000 $km^2$ in northeastern Italy \citep{dallan2023}, featuring diverse terrain with elevations ranging from up to $3{,}905$ meters above sea level and notable climatic heterogeneity.
Indeed, the mean annual precipitation is about $800mm$  per year in the south‐eastern and mostly flat part of the domain, increasing to $2300$–$2500mm$ per year toward the central part, where the Prealps form the first orographic barrier. 
The considered area is subjected to the so-called \textit{orographic enhancement}. This effect is related to the orographic lift: when the air masses encounter mountains, they are forced to rise over the obstacle on the windward side. Air cools down and, if the moisture is sufficiently large, the water vapor condenses into clouds and precipitation. On the leeward side, while air is losing most of humidity, condensation reduces and, as a consequence, also the rainfall amount decreases. Thus, this effect is related to a positive dependence between rainfall amount and altitude. However, this phenomenon is triggered if the humidity is sufficiently high. This typically occurs when air masses come from the sea and encounter the first line of mountains. Nonetheless, after few lines of mountains, most of the moisture is dropped and the effect disappears. Furthermore, orographic enhancement is influenced by a multitude of physical processes. Thus, it is hard to establish a clear linear dependence between altitude and rainfall amount. In the Alps, rainfall typically increases with altitude in the range of $800$m to $1200$m , however, over this bound the dependence may become negative. Consistently with this, fitting a linear model may fail to capture the complex relationship between altitude and rainfall making the proposed semi-parametric model an appealing alternative. 

\begin{figure}[t]
	\centering
	\includegraphics{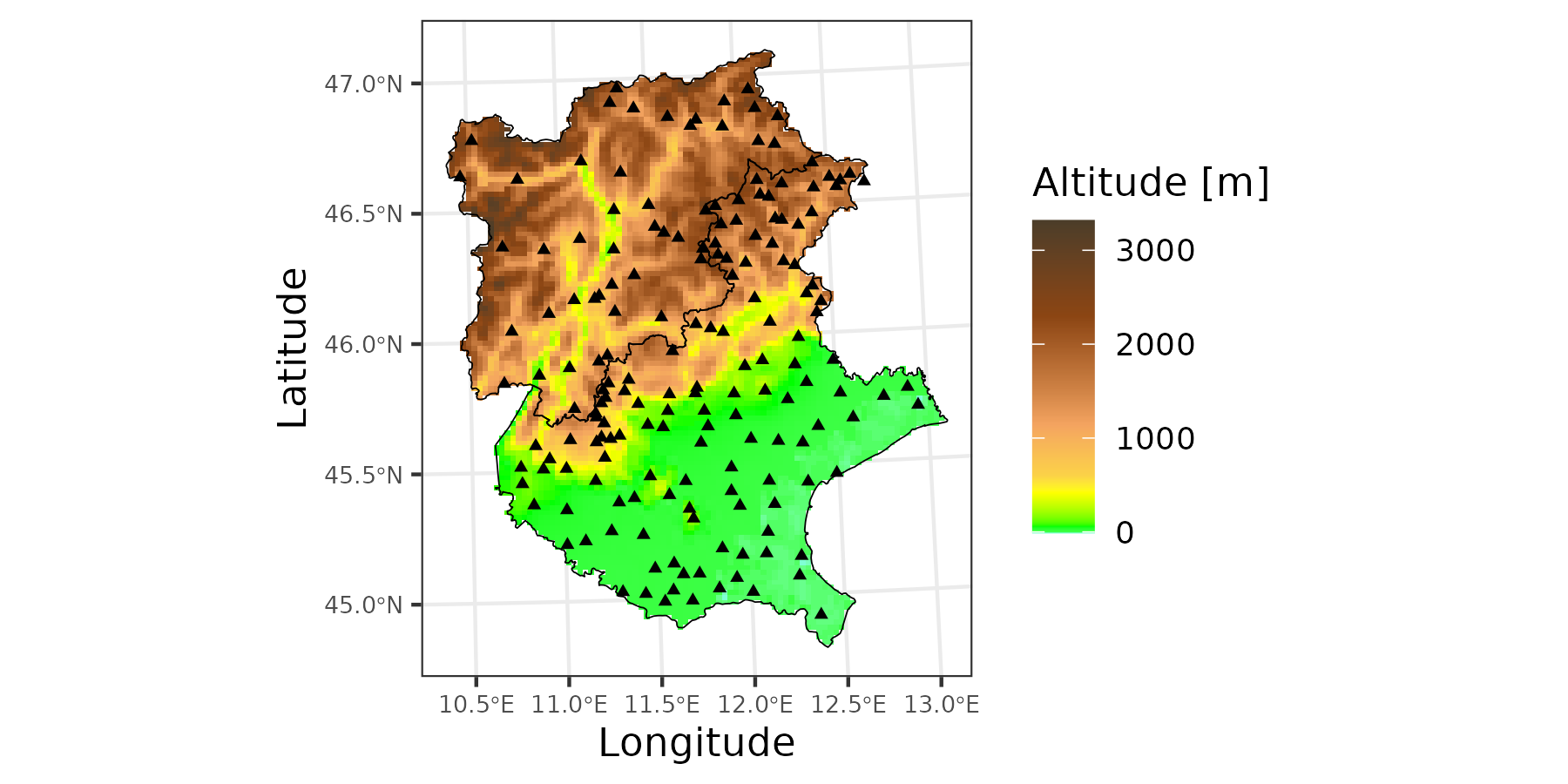}
	\caption{Northeastern Italy Data Analysis: map of orography for Veneto and Trentino-Alto Adige with locations of the weather stations (triangles).}
		\label{fig:map_altStat}
	\end{figure}

The dataset contains the hourly rainfall time series for $174$ weather station spanning from $1981$ to $2020$ and located in the two regions of Trentino-Alto- Adige and Veneto, in the northeastern Italy. For each station, the dataset contains also the altitude and the UTC coordinates in the reference system WGS84 zone 32N. Nonetheless, the series are not recorded at the same time for all stations. The maximum overlapping period is from $2000$ to $2008$, so we restrict our analysis to this time interval. We aggregate hourly data in daily rainfall events. The range of altitudes of the stations is from $-3$ to $2235$ meters above sea level. Figure \ref{fig:map_altStat} displays a map of altitude along with the locations of the weather stations.

\begin{figure}
	\centering
	\includegraphics{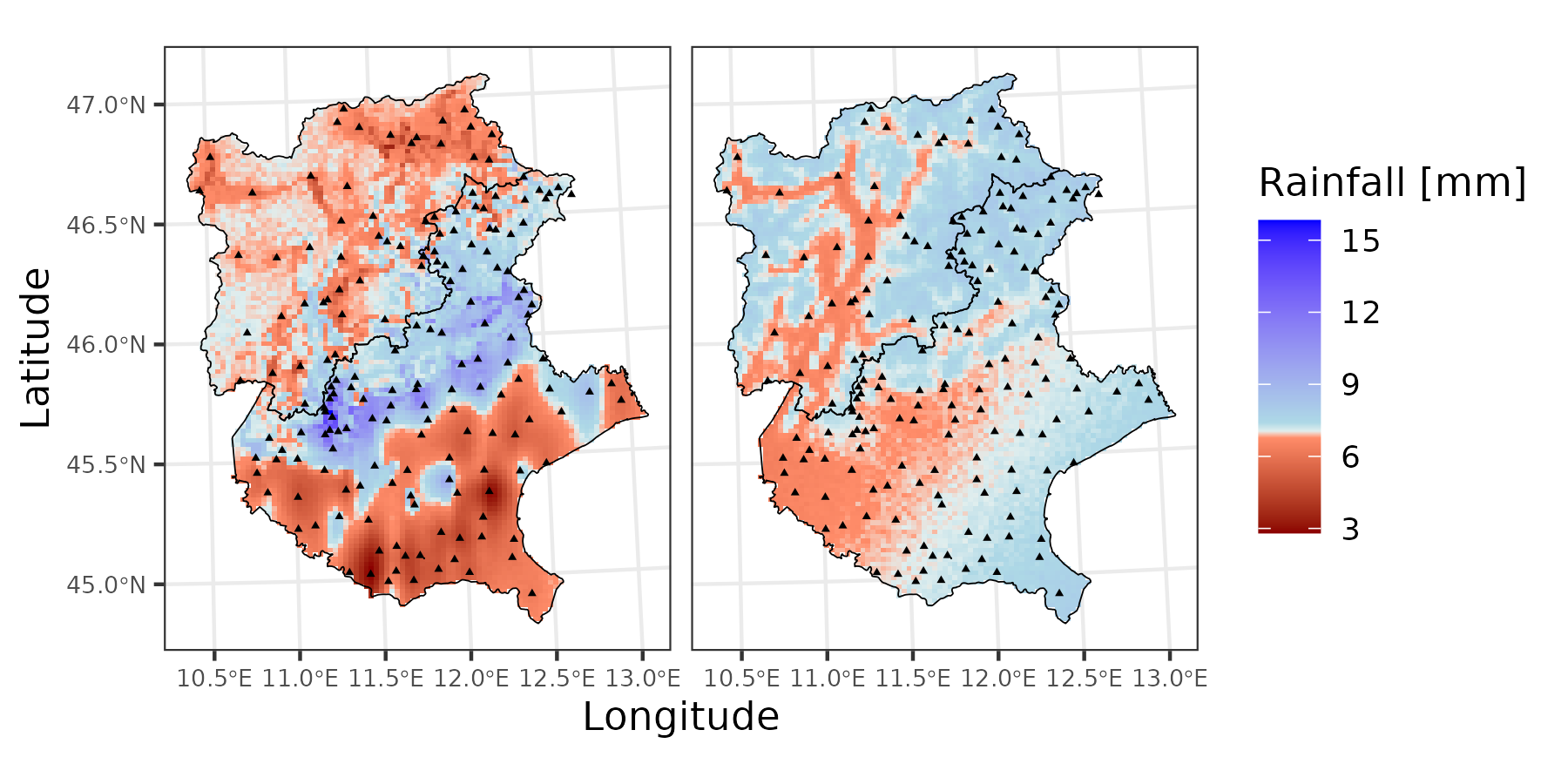}
	\caption{Northeastern Italy Data Analysis: posterior median for the Weibull expected value for a rainfall event: semi-parametric (left) vs parametric (right).}
	\label{fig:map_mean}
\end{figure}

We run the proposed MCMC algorithm using longitude, latitude, and altitude as covariates, both for the semi-parametric and parametric models. In both cases we perform $1.01 \times 10^5$ complete scans, we discard the first $1000$ as burn-in and, for memory saving purposes, we store $1$ scan every $10$. Thus, we end-up with a sample of $10^4$ posterior draws. For each generated value, we forecast the parameters of the unobserved points to obtain a posterior sample of them. For every draw of parameters, we compute the local expectation of the Weibull random variable associated to the unobserved point $s^\ast_{m^\ast} \, , \; m^\ast = 1, 2 , \dots, M^\ast$, as
\begin{equation*}
	\mathrm{E}(W_{i,j}(s^\ast_{m^\ast}) \vert \gamma_j(s^\ast_{m^\ast}), \delta_j(s^\ast_{m^\ast})) = \exp \big( \delta_j(s^\ast_{m^\ast}) \big) \, \Gamma \Big( 1 + \exp \big( -\gamma_j(s^\ast_{m^\ast}) \big) \Big).
\end{equation*}
Thus, we obtain a posterior sample of the mean rainfall amount per single event for each unobserved point; Figure \ref{fig:map_mean} displays the posterior median of this quantity for the semi-parametric and parametric models. Similarly, the expected value of rainfall amount per year is computed  as 
\begin{samepage}
	\begin{align*}
		\mathrm{E} \Bigg( \sum_{i = 1}^{N_j(s^\ast_{m^\ast})} W_{i,j}(s^\ast_{m^\ast}) \vert& \pi_j(s^\ast_{m^\ast}), \gamma_j(s^\ast_{m^\ast}), \delta_j(s^\ast_{m^\ast}) \Bigg) \\
		&= 365 \, \Big( 1 + e^{\pi_j(s^\ast_{m^\ast})} \Big) ^{-1} \, \mathrm{E} \big( W_{i,j}(s^\ast_{m^\ast}) \vert \gamma_j(s^\ast_{m^\ast}), \delta_j(s^\ast_{m^\ast}) \big).
	\end{align*}
\end{samepage}
The maps for the latter quantity is reported in the Supplementary Materials. 

Based on the findings of the simulation study, we expect that the semi-parametric model shows a more realistic pattern. In particular, it is able to capture the orographic enhancement for the first lines of mountains. As expected the parametric model is not able to do the same: it overestimates this phenomenon in the north-east of the map but underestimates it in the center-west and precisely in this area the data show the highest values of rainfall.

\begin{figure}
	\centering
	\includegraphics{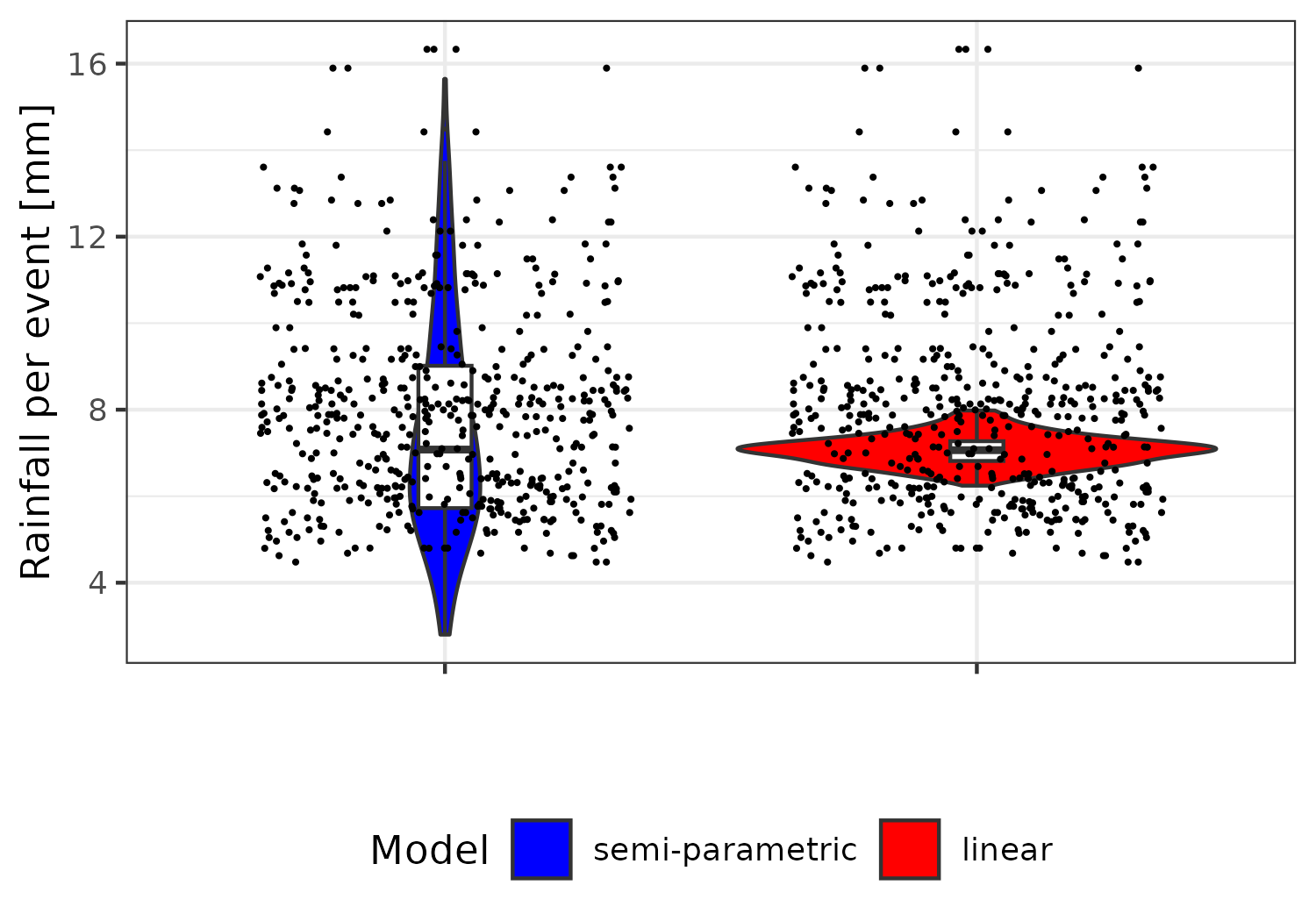}
	\caption{Northeastern Italy Data Analysis: violin plots for mean of rainfall amount per single event at the observed points: semi-parametric model (left) vs linear model (right) against jittered scatter plot of empirical values.}
	\label{fig:violin_mean}
\end{figure}

\begin{figure}
	\centering
	\includegraphics{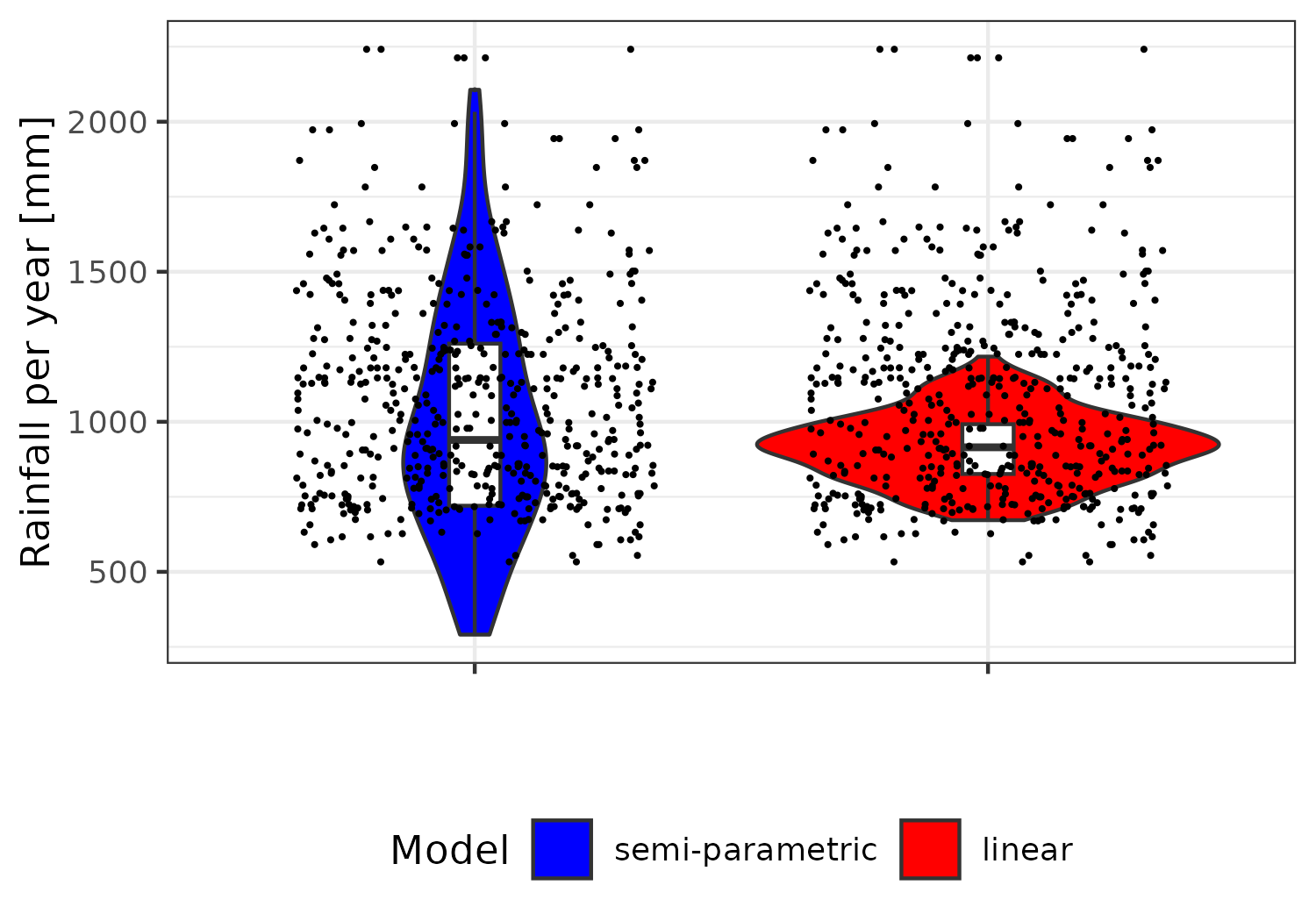}
	\caption{Northeastern Italy Data Analysis: violin plots for mean of rainfall amount per year at the observed points: semi-parametric model (left) vs linear model (right) against scatter plot of empirical values.}
	\label{fig:violin_total}
\end{figure}

The parametric model's lower accuracy is dramatically evident when comparing the empirical means of single-event magnitudes and annual rainfall amounts at observed points with those estimated by the models.
Figures \ref{fig:violin_mean} and \ref{fig:violin_total} display violin plots for the mean rainfall per event and per year, respectively, along with a jittered scatter plot of the empirical averages. The parametric model performs poorly, failing to capture the full range of observed data, while the semi-parametric model demonstrates a much better ability to reflect the data's variability.


\section{Conclusions} \label{sec:concl}

In this paper we developed a Bayesian semi-parametric model based on latent GPs for the analysis of spatially distributed rainfall data. For  posterior computation, we relied on an MCMC strategy based on elliptical slice sampling. In this way, we proposed an easy-to-use method that does not require tuning parameters to be selected by the user and it is also rejection-free. This is particularly useful in the context of nonparametric spatial data analysis, where the number of points can be very large. We showed the effectiveness of the proposed model in a simulation study and in a real data application characterized by natural complexity and interaction of spatial covariates. 

The proposed semi-parametric model outperforms the popular and simplest choice of a linear model. Our model effectively captures non-linear patterns within the data, and its goodness-of-fit improves as sample size increases. Notably, the proposed specification can serve as a foundational component for recent non-asymptotic approaches to extreme rainfall modeling \citep{marani2015, zorzetto2016, marra2019} particularly in the Bayesian context \citep{stolf2023, zorzetto2024}. These approaches rely on the assumption that ordinary rainfall events follow a Weibull distribution, building extreme value estimates without depending on the traditional generalized extreme value or generalized Pareto distributions \citep{coles2001_book}. Given that the accuracy of extreme event distribution estimation heavily hinges on the validity of assumptions about ordinary events, the performance of our model may offers a valuable enhancement in these contexts.

In the analysis of rainfall in the north-east of Italy, we provided a real example where one could expect non linear patterns. This is a region characterized by a complex orography where the relation between rainfall and altitude shows opposite trends at different locations and the results obtained by the proposed approach are consistent with these characteristics. 

\bibliography{sgp.bib}

\end{document}


\maketitle

\section{Elliptical Slice Sampling Steps} \label{sec:ellip_slice}

We report the elliptical slice sampling step for the posterior computation algorithm scheme. At the step $3$, the detailed update $(a)$ is the following:
\begin{enumerate}
	\item[(a)] The density of $\boldsymbol{\lambda_\pi}^{(k+1)} \vert \sigma^{2, (k+1)}_\pi, \boldsymbol{\mu_\pi}^{(k)}, \psi^{(k)}_\pi$ is propotional to
				\begin{align*}
					 f & \left( \boldsymbol{\mu_\pi}^{(k)} \vert \sigma^{2, (k+1)}_\pi, \boldsymbol{\lambda_\pi}^{(k+1)}, \psi^{(k)}_\pi \right) f \left( \boldsymbol{\lambda_\pi}^{(k+1)} \right) \, ,
				\end{align*}
	and the second factor is a product of log-normal densities with logarithmic mean $0$. Thus, it is possible to use an ESS step on the log-scale for updating $\boldsymbol{\lambda_\pi}^{(k+1)}$:
	\vspace{-0.5cm}
	\begin{center}
		\renewcommand{\arraystretch}{1.5}
		\begin{longtable}{l l}
				1. & \textbf{Sample} $\alpha \sim U(0, 2 \pi)$ \\
				2. & \textbf{Set} $\alpha_\mathrm{min} = \alpha - 2 \pi$ and $\alpha_\mathrm{max} = \alpha$ \\
				3. & \textbf{Sample} $u \sim U(0, 1)$, \\
				4. & \textbf{Set} $y = \log f \left( \boldsymbol{\mu_\pi}^{(k)} \vert \sigma^{2, (k+1)}_\pi, \boldsymbol{\lambda_\pi}^{(k)}, \psi^{(k)}_\pi \right) + \log u$ \\
				5. & \textbf{For} $h = 1, 2, \dots, p$: \\
					 & \quad \textbf{Sample} $\widetilde{\lambda}_{\pi, h} \sim \logn(0, 2)$ \\
				6. & \textbf{For} $h = 1, 2, \dots, p$: \\
					 & \quad \textbf{Set} $\widehat{\lambda}_{\pi, h} = \exp \left( \log \lambda^{(k)}_{\pi,h} \cos \alpha + \log \widetilde{\lambda}_{\pi, h} \sin \alpha \right)$ \\
				7. & \textbf{If} $ \log f \left( \boldsymbol{\mu_\pi}^{(k)} \vert \sigma^{2, (k+1)}_\pi, \boldsymbol{\widehat{\lambda}_\pi}, \psi^{(k)}_\pi \right) \ge y$ \\
					 & \quad \textbf{Accept} $\boldsymbol{\lambda_\pi}^{(k+1)} = \boldsymbol{\widehat{\lambda}_\pi}$ \\
				8. & \textbf{Else} \\
					 & \quad \textbf{If} $\alpha < 0$ \textbf{then} $\alpha_\mathrm{min} = \alpha$ \textbf{else} $\alpha_\mathrm{max} = \alpha$ \\
					 & \quad \textbf{Sample} $\alpha \sim U(\alpha_\mathrm{min}, \alpha_\mathrm{max})$ \\
					 & \quad \textbf{GoTo} 6.
		\end{longtable}
	\end{center}
	\vspace{-0.9cm}
	Notice that, in 4. the density of $\boldsymbol{\mu_\pi}^{(k)}$ is computed by taking the current values of the length scales $\boldsymbol{\lambda_\pi}^{(k)}$ while in 7. it is used the proposal $\boldsymbol{\widehat{\lambda}_\pi}$.
\end{enumerate}

At the step $4$, the detailed updates $(a)$ and $(b)$ are the following:
\begin{enumerate}
	\item[(a)] The density of $\psi^{(k+1)}_\pi, \boldsymbol{\mu_\pi}^{(k+1)}, \boldsymbol{\pi}^{(k+1)} \vert \boldsymbol{N}, \tau^{2, (k+1)}_\pi, \sigma^{2, (k+1)}_\pi, \boldsymbol{\lambda_\pi}^{(k+1)}, \zeta^{k+1}_{\psi, \pi}$ is propotional to
				\begin{equation*}
					 f \left( \boldsymbol{N} \vert \boldsymbol{\pi}^{(k+1)} \right) f \left( \boldsymbol{\pi}^{(k+1)}, \boldsymbol{\mu_\pi}^{(k+1)}, \psi^{(k+1)}_\pi \vert \tau^{2, (k+1)}_\pi, \sigma^{2, (k+1)}_\pi, \boldsymbol{\lambda_\pi}^{(k+1)}, \zeta^{k+1}_{\psi, \pi} \right) \, ,
				\end{equation*}
				and the second factor is a joint Gaussian density with mean equal to the null vector, thus it is possible to use an ESS step for updating $\psi^{(k+1)}_\pi, \boldsymbol{\mu_\pi}^{(k+1)}, \boldsymbol{\pi}^{(k+1)}$:
				\vspace{-0.5cm}
				\begin{center}
					\renewcommand{\arraystretch}{1.5}
					\begin{longtable}{l l}
							1. & \textbf{Sample} $\alpha \sim U(0, 2 \pi)$ \\
							2. & \textbf{Set} $\alpha_\mathrm{min} = \alpha - 2 \pi$ and $\alpha_\mathrm{max} = \alpha$ \\
							3. & \textbf{Sample} $u \sim U(0, 1)$, \\
							4. & \textbf{Set} $y = \log f \left( \boldsymbol{N} \vert \boldsymbol{\pi}^{(k)} \right) + \log u$ \\
							5. & \textbf{Sample} $\widetilde{\psi}_\pi \sim N \left( 0, \zeta^{(k+1)}_{\psi, \pi} \right)$ \\
							6. & \textbf{Sample} $\boldsymbol{\widetilde{\mu}_\pi} \sim N \left( \widetilde{\psi}_\pi 1_M, K_\pi(\boldsymbol{s}, \boldsymbol{s}^\prime \vert \sigma_\pi^{2, (k+1)}, \boldsymbol{\lambda_\pi}^{(k+1)}) \right)$ \\
							7. & \textbf{For} $m = 1, 2, \dots, M$: \\
								 & \quad \textbf{For} $j = 1, 2, \dots, T$ \\
								 & \quad \quad \textbf{Sample} $\widetilde{\pi}_j(s_m) \sim N \left( \widetilde{\mu}_\pi(s_m), \tau^{2, k+1}_\pi \right)$ \\
							8. & \textbf{Set} $\widehat{\psi}_\pi = \psi^{(k)}_\pi \cos \alpha + \widetilde{\psi}_\pi \sin \alpha$ \\
							9. & \textbf{Set} $\boldsymbol{\widehat{\mu}_\pi} = \boldsymbol{\mu_\pi}^{(k)} \cos \alpha + \boldsymbol{\widetilde{\mu}_\pi} \sin \alpha$ \\
							10. & \textbf{Set} $\boldsymbol{\widehat{\pi}} = \boldsymbol{\pi}^{(k)} \cos \alpha + \boldsymbol{\widetilde{\pi}} \sin \alpha$ \\
							11. & \textbf{If} $ \log f \left( \boldsymbol{N} \vert \boldsymbol{\widehat{\pi}} \right) \ge y$ \\
								 & \quad \textbf{Accept} $\psi^{(k+1)}_\pi = \widehat{\psi}_\pi$, $\boldsymbol{\mu_\pi}^{(k+1)} = \boldsymbol{\widehat{\mu}_\pi}$, $\boldsymbol{\pi}^{(k+1)} = \boldsymbol{\widehat{\pi}}$ \\
							12. & \textbf{Else} \\
								 & \quad \textbf{If} $\alpha < 0$ \textbf{then} $\alpha_\mathrm{min} = \alpha$ \textbf{else} $\alpha_\mathrm{max} = \alpha$ \\
								 & \quad \textbf{Sample} $\alpha \sim U(\alpha_\mathrm{min}, \alpha_\mathrm{max})$ \\
								 & \quad \textbf{GoTo} 8.
					\end{longtable}
			\end{center}
			\vspace{-0.9cm}
			Notice that, in 4. the density of $\boldsymbol{N}$ is computed by taking the current values of the binomial parameters $\boldsymbol{\pi}^{(k)}$ while in 11. it is used the proposal $\boldsymbol{\widehat{\pi}}$.
	
	\vspace{2.5mm}
	\item[(b)] The density of $\psi^{(k+1)}_\gamma, \boldsymbol{\mu_\gamma}^{(k+1)}, \boldsymbol{\gamma}^{(k+1)}, \psi^{(k+1)}_\delta, \boldsymbol{\mu_\delta}^{(k+1)}, \boldsymbol{\delta}^{(k+1)} \vert \boldsymbol{W}, \boldsymbol{N}, $ $ \tau^{2, (k+1)}_\gamma, \sigma^{2, (k+1)}_\gamma, \boldsymbol{\lambda_\gamma}^{(k+1)}, \zeta^{k+1}_{\psi, \gamma}, \tau^{2, (k+1)}_\delta, \sigma^{2, (k+1)}_\delta, \boldsymbol{\lambda_\delta}^{(k+1)}, \zeta^{k+1}_{\psi, \delta}$ is proportional to
				\begin{align*}
					&f \left( \boldsymbol{W} \vert \boldsymbol{N}, \boldsymbol{\gamma}^{(k+1)}, \boldsymbol{\delta}^{(k+1)} \right) \times \\
					&\times f \left( \boldsymbol{\gamma}^{(k+1)}, \boldsymbol{\mu_\gamma}^{(k+1)}, \psi^{(k+1)}_\gamma \vert \tau^{2, (k+1)}_\gamma, \sigma^{2, (k+1)}_\gamma, \boldsymbol{\lambda_\gamma}^{(k+1)}, \zeta^{k+1}_{\psi, \gamma} \right) \times \\
					&\times f \left( \boldsymbol{\delta}^{(k+1)}, \boldsymbol{\mu_\delta}^{(k+1)}, \psi^{(k+1)}_\delta \vert \tau^{2, (k+1)}_\delta, \sigma^{2, (k+1)}_\delta, \boldsymbol{\lambda_\delta}^{(k+1)}, \zeta^{k+1}_{\psi, \delta} \right) \, ,
				\end{align*}
	and the latter $2$ factors are a product between $2$ joint Gaussian densities with both means equal to the null vector, thus it is possible to use an ESS step for updating $\psi^{(k+1)}_\gamma, \boldsymbol{\mu_\gamma}^{(k+1)}, \boldsymbol{\gamma}^{(k+1)}$ and $\psi^{(k+1)}_\delta, \boldsymbol{\mu_\delta}^{(k+1)}, \boldsymbol{\delta}^{(k+1)}$:
	\vspace{-0.5cm}
	\begin{center}
		\renewcommand{\arraystretch}{1.5}
		\begin{longtable}{l l}
				1. & \textbf{Sample} $\alpha \sim U(0, 2 \pi)$ \\
				2. & \textbf{Set} $\alpha_\mathrm{min} = \alpha - 2 \pi$ and $\alpha_\mathrm{max} = \alpha$ \\
				3. & \textbf{Sample} $u \sim U(0, 1)$, \\
				4. & \textbf{Set} $y = \log f \left( \boldsymbol{W} \vert \boldsymbol{N}, \boldsymbol{\gamma}^{(k)}, \boldsymbol{\delta}^{(k)} \right) + \log u$ \\
				5. & \textbf{Sample} $\widetilde{\psi}_\gamma \sim N \left( 0, \zeta^{(k+1)}_{\psi, \gamma} \right)$ \\
				6. & \textbf{Sample} $\boldsymbol{\widetilde{\mu}_\gamma} \sim N \left( \widetilde{\psi}_\gamma 1_M, K_\gamma(\boldsymbol{s}, \boldsymbol{s}^\prime \vert \sigma_\gamma^{2, (k+1)}, \boldsymbol{\lambda_\gamma}^{(k+1)}) \right)$ \\
				7. & \textbf{For} $m = 1, 2, \dots, M$: \\
					 & \quad \textbf{For} $j = 1, 2, \dots, T$ \\
					 & \quad \quad \textbf{Sample} $\widetilde{\gamma}_j(s_m) \sim N \left( \widetilde{\mu}_\gamma(s_m), \tau^{2, k+1}_\gamma \right)$ \\
				8. & \textbf{Sample} $\widetilde{\psi}_\delta, \boldsymbol{\widetilde{\mu}_\delta}, \boldsymbol{\widetilde{\delta}}$ similarly to 5. -- 7. \\
				9. & \textbf{Set} $\widehat{\psi}_\gamma = \psi^{(k)}_\gamma \cos \alpha + \widetilde{\psi}_\gamma \sin \alpha$ \\
				10. & \textbf{Set} $\boldsymbol{\widehat{\mu}_\gamma} = \boldsymbol{\mu_\gamma}^{(k)} \cos \alpha + \boldsymbol{\widetilde{\mu}_\gamma} \sin \alpha$ \\
				11. & \textbf{Set} $\boldsymbol{\widehat{\gamma}} = \boldsymbol{\gamma}^{(k)} \cos \alpha + \boldsymbol{\widetilde{\gamma}} \sin \alpha$ \\
				12. & \textbf{Set} $\widehat{\psi}_\delta, \boldsymbol{\widehat{\mu}_\delta}, \boldsymbol{\widehat{\delta}}$ similarly to 9. -- 11. \\
				13. & \textbf{If} $ \log f \left( \boldsymbol{W} \vert \boldsymbol{N}, \boldsymbol{\widehat{\gamma}}, \boldsymbol{\widehat{\delta}} \right) \ge y$ \\
					 & \quad \textbf{Accept} $\psi^{(k+1)}_\gamma = \widehat{\psi}_\gamma$, $\boldsymbol{\mu_\gamma}^{(k+1)} = \boldsymbol{\widehat{\mu}_\gamma}$, $\boldsymbol{\gamma}^{(k+1)} = \boldsymbol{\widehat{\gamma}}$ \\
					 & \quad \textbf{Accept} $\psi^{(k+1)}_\delta = \widehat{\psi}_\delta$, $\boldsymbol{\mu_\delta}^{(k+1)} = \boldsymbol{\widehat{\mu}_\delta}$, $\boldsymbol{\delta}^{(k+1)} = \boldsymbol{\widehat{\delta}}$ \\
				14. & \textbf{Else} \\
					 & \quad \textbf{If} $\alpha < 0$ \textbf{then} $\alpha_\mathrm{min} = \alpha$ \textbf{else} $\alpha_\mathrm{max} = \alpha$ \\
					 & \quad \textbf{Sample} $\alpha \sim U(\alpha_\mathrm{min}, \alpha_\mathrm{max})$ \\
					 & \quad \textbf{GoTo} 9.
		\end{longtable}
	\end{center}
	\vspace{-0.9cm}
	Notice that, in 4. the density of $\boldsymbol{W}$ is computed by taking the current values of the Weibull parameters $\boldsymbol{\gamma}^{(k)}$ and $\boldsymbol{\delta}^{(k)}$ while in 13. the proposals $\boldsymbol{\widehat{\gamma}}$ and $\boldsymbol{\widehat{\delta}}$ are used.
\end{enumerate}

\section{Parametric Model} \label{sec:parametric}

We report the details of the parametric linear model used as competitor. As in the semi-parametric case, the regressors are spatial coordinates and topographical features. We assume:
\begin{align*}
	\pi_j(s) &= \beta_{\pi, 0} + \sum_{h=1}^{p} \beta_{\pi, h} s_{j, h} + \eps_{\pi, j}(s) \, , \: \eps_{j, \pi}(s) \overset{i.i.d.}{\sim} N(0, \tau^2_\pi) \, , \\
	\gamma_j(s) &= \beta_{\gamma, 0} + \sum_{h=1}^{p} \beta_{\gamma, h} s_{j, h} + \eps_{\gamma, j}(s) \, , \: \eps_{j, \gamma}(s) \overset{i.i.d.}{\sim} N(0, \tau^2_\gamma) \, , \\
	\delta_j(s) &= \beta_{\delta, 0} + \sum_{h=1}^{p} \beta_{\delta, h} s_{j, h} + \eps_{\delta, j}(s) \, , \: \eps_{j, \delta}(s) \overset{i.i.d.}{\sim} N(0, \tau^2_\delta) \, .
\end{align*}
The priors for the variance of the white noises are the same as the ones used for the semi-parametric model. The priors for the coefficients are independent $\stu_2$.

From a computational perspective, we use the same approach of the semi-parametric case for the parametric model. 

Let $\zeta_{h, \pi}, \zeta_{h, \gamma}, \zeta_{h, \delta}$ be the scaling variables of $\beta_{h, \pi}, \beta_{h, \gamma}, \beta_{h, \delta}$ respectively. Let $\boldsymbol{\zeta_{\beta, \pi}} = \{ \zeta_{h, \beta, \pi}: h = 0, 1, \dots, p\}$ and let $\boldsymbol{\beta_\pi} = \{ \beta_{h,\pi}: h = 0, 1, \dots, p \}$ and let the same be for $\boldsymbol{\zeta_{\beta, \gamma}}, \boldsymbol{\beta_\delta}$ and $\boldsymbol{\zeta_{\beta, \delta}}, \boldsymbol{\beta_\delta}$.  Therefore, the updating scheme simplifies to the following steps:
\begin{enumerate}

	\item Update scaling variable of stochastic representations of Student-$t$ and inverse compounded gamma distributions:
				\vspace{2.5mm}
				\begin{enumerate}
					\item For $h = 0, 1, \dots, p$, sample $\zeta^{(k+1)}_{h, \beta, \pi} \vert \beta^{(k)}_{h,\pi}$ from
					\begin{equation*}
						\igamma \left( \frac{3}{2}, 1 + \frac{\beta^{2,(k)}_{h,\pi}}{2} \right) \, ,
					\end{equation*} 
					
					\item Sample $\zeta^{(k+1)}_{\tau^2, \pi} \vert \tau^{2, (k)}_\pi$ from
								\begin{equation*}
									\dgamma \left( \frac{5}{2}, \frac{1}{2} + \frac{1}{\tau^{2, (k)}_\pi} \right) \, ,
								\end{equation*}
					
					\item Do the same for $\boldsymbol{\zeta_{\beta, \gamma}}^{(k+1)}, \zeta^{(k+1)}_{\tau^2, \gamma}$ and $\boldsymbol{\zeta_{\beta, \delta}}^{(k+1)}, \zeta^{(k+1)}_{\tau^2, \delta}$.
				\end{enumerate}
	
	\vspace{2.5mm}
	\item Update linear regression coefficients and parameters:
	\vspace{2.5mm}
	\begin{enumerate}
		\item The density of $\boldsymbol{\beta_\pi}^{(k+1)}, \boldsymbol{\pi}^{(k+1)} \vert \boldsymbol{N}, \tau^{2, (k+1)}_\pi, \boldsymbol{\zeta_{\beta, \pi}}^{(k+1)}$ is propotional to
					\begin{equation*}
						 f \left( \boldsymbol{N} \vert \boldsymbol{\pi}^{(k+1)} \right) f \left( \boldsymbol{\pi}^{(k+1)}, \boldsymbol{\beta_\pi}^{(k+1)} \vert \tau^{2, (k+1)}_\pi, \boldsymbol{\zeta_{\beta, \pi}}^{(k+1)} \right) \, ,
					\end{equation*}
					and the second factor is joint Gaussian density with mean equal to the null vector, thus it is possible to use an ESS step for updating $\boldsymbol{\beta_\pi}^{(k+1)}, \boldsymbol{\pi}^{(k+1)}$.
		
		\vspace{2.5mm}
		\item The density of $\boldsymbol{\beta_\gamma}^{(k+1)}, \boldsymbol{\gamma}^{(k+1)}, \boldsymbol{\beta_\delta}^{(k+1)}, \boldsymbol{\delta}^{(k+1)} \vert \boldsymbol{W}, \boldsymbol{N}, \tau^{2, (k+1)}_\gamma, \boldsymbol{\zeta_{\beta, \gamma}}^{(k+1)}, \tau^{2, (k+1)}_\delta, \boldsymbol{\zeta_{\beta, \delta}}^{(k+1)}$ is proportional to
					\begin{align*}
						&f \left( \boldsymbol{W} \vert \boldsymbol{N}, \boldsymbol{\gamma}^{(k+1)}, \boldsymbol{\delta}^{(k+1)} \right)
						f \left( \boldsymbol{\gamma}^{(k+1)}, \boldsymbol{\beta_\gamma}^{(k+1)} \vert \tau^{2, (k+1)}_\gamma, \boldsymbol{\zeta_{\beta, \gamma}}^{(k+1)} \right) \times \\
						&\times f \left( \boldsymbol{\delta}^{(k+1)}, \boldsymbol{\beta_\delta}^{(k+1)} \vert \tau^{2, (k+1)}_\delta, \boldsymbol{\zeta_{\beta, \delta}}^{(k+1)} \right) \, ,
					\end{align*}
		and the latter $2$ factors are a product between $2$ joint Gaussian densities with both means equal to the null vector, thus it is possible to use an ESS step for updating $\boldsymbol{\beta_\gamma}^{(k+1)}, \boldsymbol{\gamma}^{(k+1)}$ and $\boldsymbol{\beta_\delta}^{(k+1)}, \boldsymbol{\delta}^{(k+1)}$.
	\end{enumerate}
	 
\end{enumerate}
Also the parametric model allow us to forecast the number of events and their magnitude at unobserved points. Indeed, given the values of the linear regression coefficients and the variance of white noises at the end of $k+1$ iteration, it is straightforward to use the linear model in order to forecast the parameters of binomial and Weibull distributions at unobserved points. Also for the case of parametric model we use, as Bayesian point estimator, the posterior median.

\section{Simulation Study: Details} \label{sec:simulation_details}

\begin{figure}
	\centering
	\includegraphics{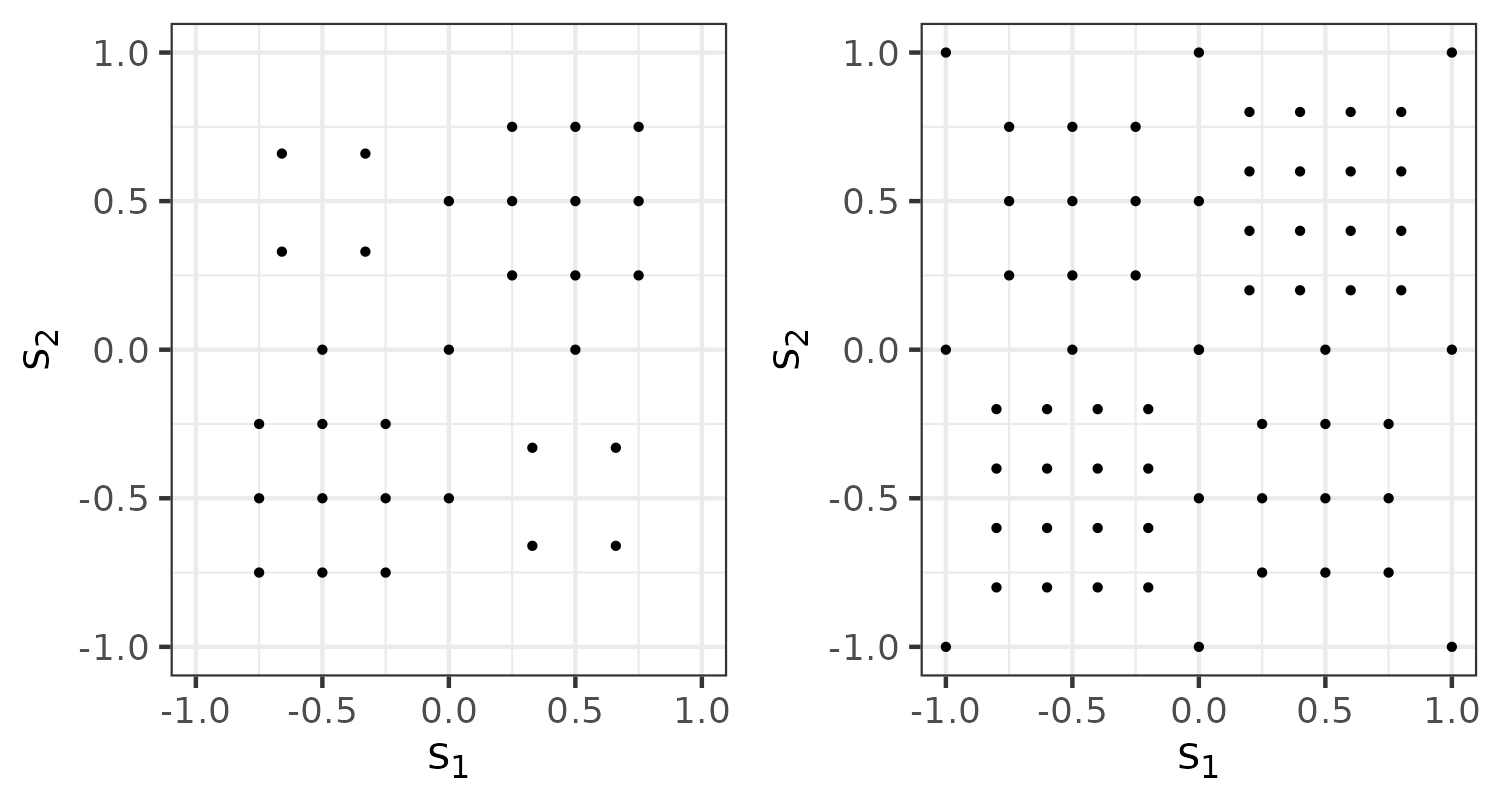}
	\caption{Simulation study: observed point locations for different sample sizes: small (left) and large (right).}
	\label{fig:point_locs_both}
\end{figure}

Figure \ref{fig:point_locs_both} displays the observed point locations for small and large sample size cases.
Regarding the expressions generating the true parameter values $\gamma^\dag(s)$ and $\delta^\dag(s)$, for the non-linear scenario we set
\begin{align*}
	\gamma^\dag(s_m) &= \frac{69}{100} + \frac{52}{25} \left( \frac{2 e^{ 2 s_{m,1} - 1 - \vert 2 s_{m,2} - 1 \vert }}{\left( 1 + e^{ 2 s_{m,1} - 1 } \right)^2} - 
	\sqrt{\frac{8}{\pi}} \frac{e^{-(2 s_{m,2} + 1)^2}}{1 + (2 s_{1,m} + 1)^2}\right) \, , \\
	\delta^\dag(s) &= \sin(\pi s_{m,1}) + \sin(\pi s_{m,2}) \, .
\end{align*}
while, in the linear scenario, we set
\begin{align*}
	\gamma^\dag(s_m) &= \frac{17}{25} + \frac{29}{100} s_{m,1} + \frac{7}{10} s_{m,2} \, , \\ 
	\delta^\dag(s) &= \frac{7}{5} s_{m,1} + \frac{58}{100} s_{m,2} \, .
\end{align*}
Notice that, for the shape parameters in the non-linear case, the function is obtained by applying an affine transformation in a combination of Cauchy, Laplace, Gaussian, and logistic densities. In particular, we have an hole in $(-0.5, -0.5)$ and a peak in $(0.5, 0.5)$ but the shapes of the $2$ curves are different. The linear case is a simple linear combination of the spatial coordinates. The coefficients are chosen in order to have roughly the same range of the parameters with respect to the non-linear case.

\begin{figure}
	\centering
	\includegraphics{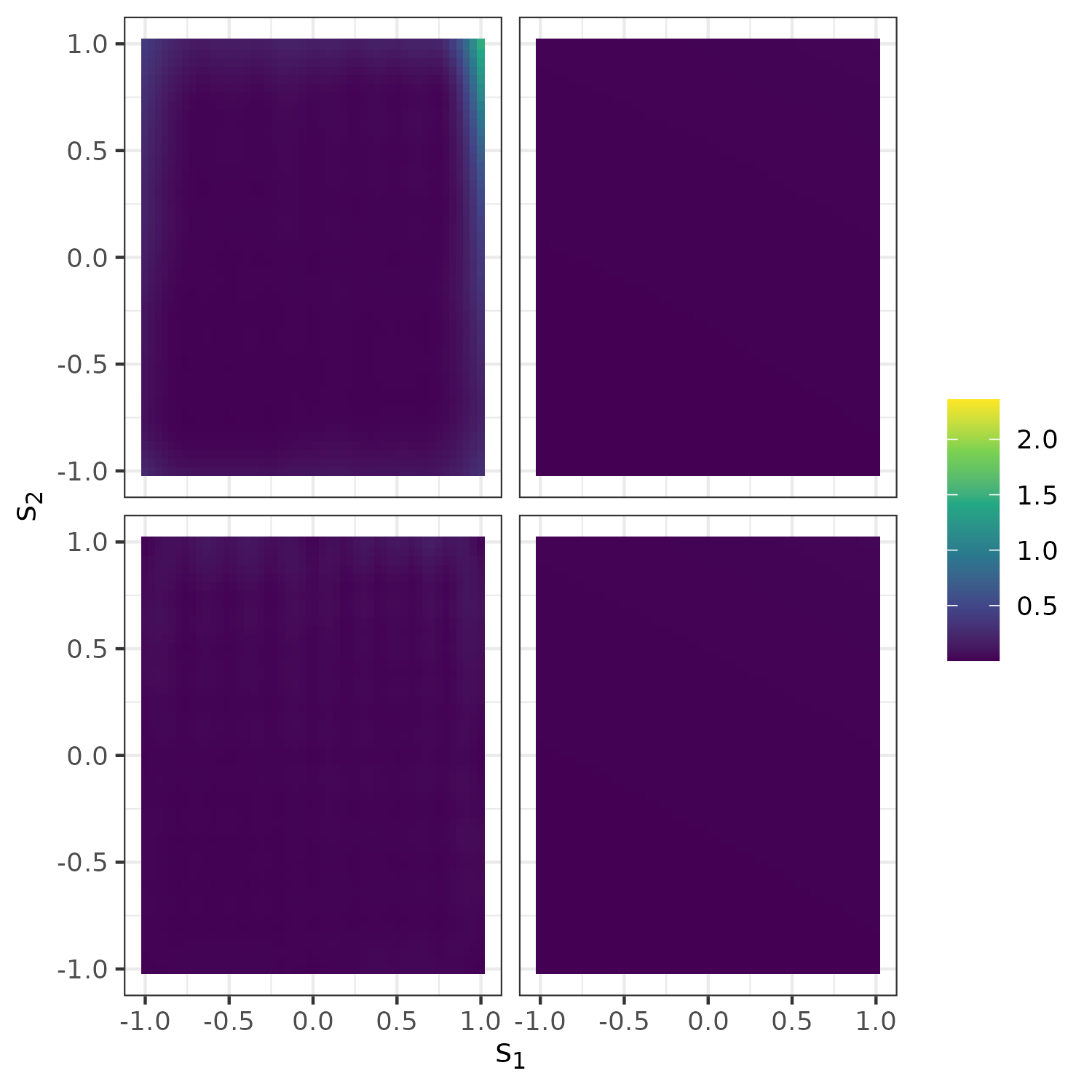}
	\caption{Simulation study: logarithm of sample mean of Kullback-Leibler divergence estimates for the linear scenario: semi-parametric (left) vs parametric (right), small sample size (top) and large sample size (bottom).}
	\label{fig:klmaps_lm}
\end{figure}

Figure \ref{fig:klmaps_lm} reports the logarithm of the average posterior medians of the Kullback-Leibler divergences, obtained averaging over the $R$ replicates, for each pixel of the map in the linear scenario.

\section{Analysis of the Rainfall in the North-East of Italy: Details} 

\begin{figure}
	\centering
	\includegraphics{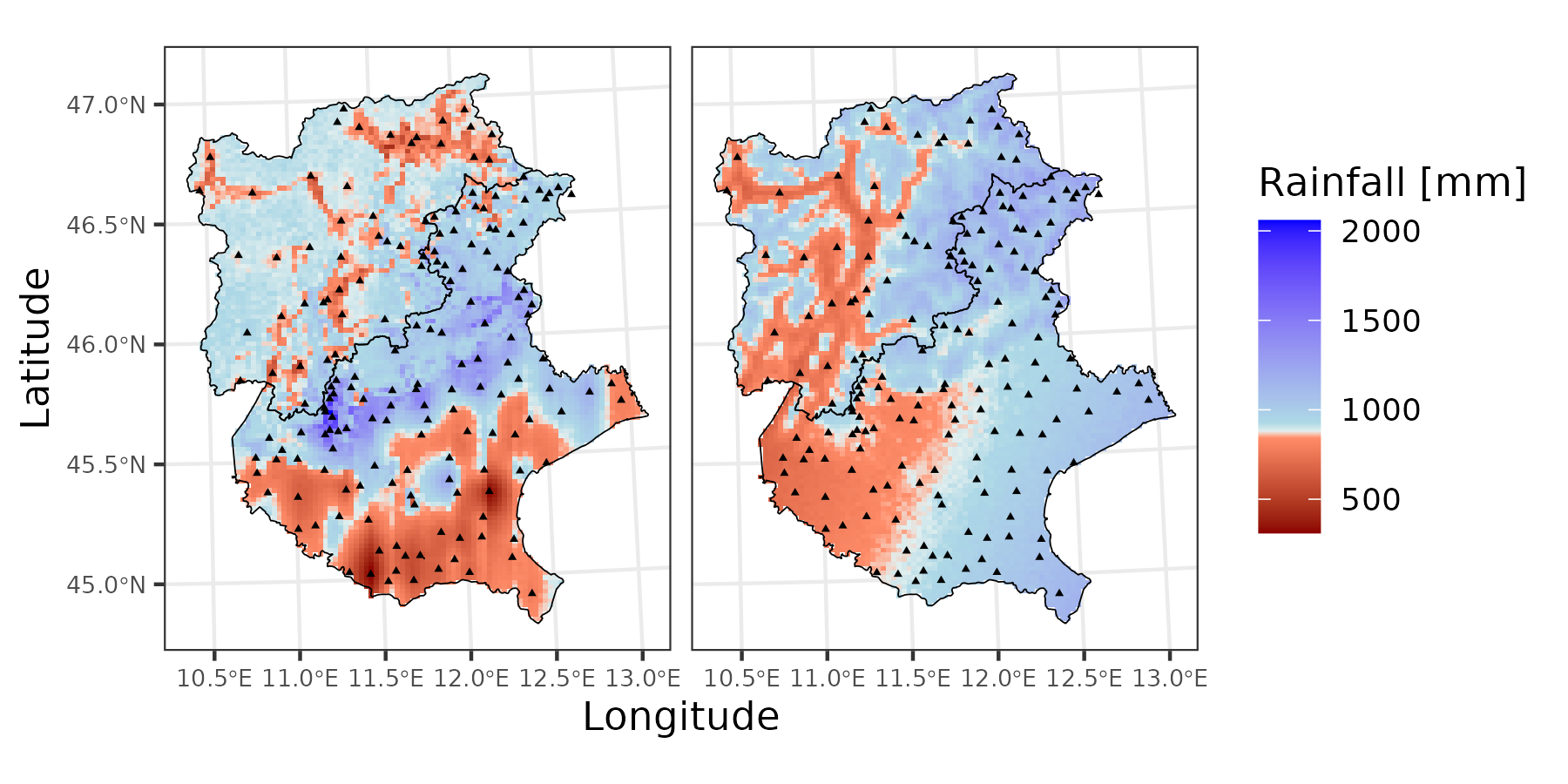}
	\caption{Northeastern Italy Data Analysis: expected value of rainfall amount per year for the semi-parametric (left) and parametric (right) models.}
	\label{fig:map_total}
\end{figure}

%% file: figures/dag.tex
	\begin{tikzpicture}
		\Vertex[x=2*5,y=9,color=gray, size=0.1,position=above, label=$\zeta_{\psi, \pi}$,fontsize=\large]{zeta_psi_pi}
		\Vertex[x=2*5,y=8,color=gray, size=0.1,position=above, label=$\zeta_{\sigma^2, \pi}$,fontsize=\large]{zeta_sigmaSq_pi}
		\Vertex[x=2*5,y=6,color=gray, size=0.1,position=above, label=$\zeta_{\tau^2, \pi}$,fontsize=\large]{zeta_tauSq_pi}
		\Vertex[x=2*5,y=4,color=gray, size=0.1,position=above, label=$\zeta_{\psi, \gamma}$,fontsize=\large]{zeta_psi_gamma}
		\Vertex[x=2*5,y=3,color=gray, size=0.1,position=above, label=$\zeta_{\sigma^2, \gamma}$,fontsize=\large]{zeta_sigmaSq_gamma}
		\Vertex[x=2*5,y=1,color=gray, size=0.1,position=above, label=$\zeta_{\tau^2, \gamma}$,fontsize=\large]{zeta_tauSq_gamma}
		\Vertex[x=2*5,y=-1,color=gray, size=0.1,position=above, label=$\zeta_{\psi, \delta}$,fontsize=\large]{zeta_psi_delta}
		\Vertex[x=2*5,y=-2,color=gray, size=0.1,position=above, label=$\zeta_{\sigma^2, \delta}$,fontsize=\large]{zeta_sigmaSq_delta}
		\Vertex[x=2*5,y=-4,color=gray, size=0.1,position=above, label=$\zeta_{\tau^2, \delta}$,fontsize=\large]{zeta_tauSq_delta}
		\Vertex[x=2*4,y=9,color=gray, size=0.1,position=above, label=$\psi_\pi$,fontsize=\large]{psi_pi}
		\Vertex[x=2*4,y=8,color=gray, size=0.1,position=above, label=$\sigma^2_\pi$,fontsize=\large]{sigmaSq_pi}
		\Vertex[x=2*4,y=7,color=gray, size=0.1,position=above, label=$\boldsymbol{\lambda}_{\pi}$,fontsize=\large]{lambda_pi}
		\Vertex[x=2*4,y=4,color=gray, size=0.1,position=above, label=$\psi_\gamma$,fontsize=\large]{psi_gamma}
		\Vertex[x=2*4,y=3,color=gray, size=0.1,position=above, label=$\sigma^2_\gamma$,fontsize=\large]{sigmaSq_gamma}
		\Vertex[x=2*4,y=2,color=gray, size=0.1,position=above, label=$\boldsymbol{\lambda}_{\gamma}$,fontsize=\large]{lambda_gamma}
		\Vertex[x=2*4,y=-1,color=gray, size=0.1,position=above, label=$\psi_\delta$,fontsize=\large]{psi_delta}
		\Vertex[x=2*4,y=-2,color=gray, size=0.1,position=above, label=$\sigma^2_\delta$,fontsize=\large]{sigmaSq_delta}
		\Vertex[x=2*4,y=-3,color=gray, size=0.1,position=above, label=$\boldsymbol{\lambda}_{\delta}$,fontsize=\large]{lambda_delta}
		\Vertex[x=2*3,y=8,color=gray, size=0.1,position=above left, label=$\boldsymbol{\mu_\pi}$,fontsize=\large]{mu_pi}
		\Vertex[x=2*3,y=6,color=gray, size=0.1,position=above, label=$\tau^2_\pi$,fontsize=\large]{tauSq_pi}
		\Vertex[x=2*3,y=3,color=gray, size=0.1,position=above left, label=$\boldsymbol{\mu_\gamma}$,fontsize=\large]{mu_gamma}
		\Vertex[x=2*3,y=1,color=gray, size=0.1,position=above, label=$\tau^2_\gamma$,fontsize=\large]{tauSq_gamma}
		\Vertex[x=2*3,y=-2,color=gray, size=0.1,position=above left, label=$\boldsymbol{\mu_\delta}$,fontsize=\large]{mu_delta}
		\Vertex[x=2*3,y=-4,color=gray, size=0.1,position=above, label=$\tau^2_\delta$,fontsize=\large]{tauSq_delta}
		\Vertex[x=2*2,y=7,color=gray, size=0.1,position=above left, label=$\boldsymbol{\pi}$,fontsize=\large]{pi}
		\Vertex[x=2*2,y=2,color=gray, size=0.1,position=above left, label=$\boldsymbol{\gamma}$,fontsize=\large]{gamma}
		\Vertex[x=2*2,y=-3,color=gray, size=0.1,position=below left, label=$\boldsymbol{\delta}$,fontsize=\large]{delta}
		\Vertex[x=2*1,y=7,color=gray, size=0.1,position=above left, label=$\boldsymbol{N}$,fontsize=\large]{N}
		\Vertex[x=2*1,y=-0.5,color=gray, size=0.1,position=above left, label=$\boldsymbol{W}$,fontsize=\large]{W}
		\Edge[lw=1pt,Direct](zeta_psi_pi)(psi_pi)
		\Edge[lw=1pt,Direct](zeta_sigmaSq_pi)(sigmaSq_pi)
		\Edge[lw=1pt,Direct](zeta_tauSq_pi)(tauSq_pi)
		\Edge[lw=1pt,Direct](zeta_psi_gamma)(psi_gamma)
		\Edge[lw=1pt,Direct](zeta_sigmaSq_gamma)(sigmaSq_gamma)
		\Edge[lw=1pt,Direct](zeta_tauSq_gamma)(tauSq_gamma)
		\Edge[lw=1pt,Direct](zeta_psi_delta)(psi_delta)
		\Edge[lw=1pt,Direct](zeta_sigmaSq_delta)(sigmaSq_delta)
		\Edge[lw=1pt,Direct](zeta_tauSq_delta)(tauSq_delta)
		\Edge[lw=1pt,Direct](psi_pi)(mu_pi)
		\Edge[lw=1pt,Direct](sigmaSq_pi)(mu_pi)
		\Edge[lw=1pt,Direct](lambda_pi)(mu_pi)
		\Edge[lw=1pt,Direct](psi_gamma)(mu_gamma)
		\Edge[lw=1pt,Direct](sigmaSq_gamma)(mu_gamma)
		\Edge[lw=1pt,Direct](lambda_gamma)(mu_gamma)
		\Edge[lw=1pt,Direct](psi_delta)(mu_delta)
		\Edge[lw=1pt,Direct](sigmaSq_delta)(mu_delta)
		\Edge[lw=1pt,Direct](lambda_delta)(mu_delta)
		\Edge[lw=1pt,Direct](mu_pi)(pi)
		\Edge[lw=1pt,Direct](tauSq_pi)(pi)
		\Edge[lw=1pt,Direct](mu_gamma)(gamma)
		\Edge[lw=1pt,Direct](tauSq_gamma)(gamma)
		\Edge[lw=1pt,Direct](mu_delta)(delta)
		\Edge[lw=1pt,Direct](tauSq_delta)(delta)
		\Edge[lw=1pt,Direct](pi)(N)
		\Edge[lw=1pt,Direct](gamma)(W)
		\Edge[lw=1pt,Direct](delta)(W)
		\Edge[lw=1pt,Direct](N)(W)
	\end{tikzpicture}